\documentclass[twocolumn]{aa}  

\usepackage{graphicx}
\usepackage[varg]{txfonts}

\usepackage{ulem}

\usepackage[utf8]{inputenc}

\usepackage{longtable}
\usepackage{lscape}

\usepackage{natbib,twoopt}

\begin{document}

  \title{High-resolution, 3D radiative transfer modeling: I. The grand-design spiral galaxy M\,51}

\author{Ilse De Looze\inst{1,2}
 \and Jacopo Fritz\inst{1}
 \and Maarten Baes\inst{1}
  \and George J. Bendo\inst{3}
      \and Luca Cortese\inst{4}   
      \and M{\'e}d{\'e}ric Boquien\inst{2}
      \and Alessandro Boselli\inst{5}
    \and Peter Camps\inst{1}
      \and Asantha Cooray\inst{6,7}
        \and Diane Cormier\inst{8}
        \and Jon I. Davies\inst{9}
        \and Gert De Geyter\inst{1}        
        \and Thomas M. Hughes\inst{1}
        \and Anthony P. Jones\inst{10,11}
      \and Oskar {\L}. Karczewski\inst{12}
      \and Vianney Lebouteiller\inst{13}
      \and Nanyao Lu\inst{14}
        \and Suzanne C. Madden\inst{13} 
        \and Aur{\'e}lie R{\'e}my-Ruyer\inst{10,11,13}
        \and Luigi Spinoglio\inst{15}
        \and Matthew W.L. Smith\inst{8}
        \and Sebastien Viaene\inst{1} 
        \and Christine D. Wilson\inst{16}
  } 

\institute{$^{1}$ Sterrenkundig Observatorium, Universiteit Gent, Krijgslaan 281 S9, B-9000 Gent, Belgium \\
$^{2}$ Institute of Astronomy, University of Cambridge, Madingley Road, Cambridge, CB3 0HA, UK \\
$^{3}$ UK ALMA Regional Centre Node, Jodrell Bank Centre for Astrophysics, School of Physics and Astronomy, University of Manchester,
Oxford Road, Manchester M13 9PL, United Kingdom \\
$^{4}$ Centre for Astrophysics \& Supercomputing, Swinburne University of Technology, Mail H30 - PO Box 218, Hawthorn, VIC 3122, Australia \\
$^{5}$ Laboratoire d$'$Astrophysique de Marseille - LAM, Universit{\'e} d$'$Aix-Marseille \& CNRS, UMR7326, 38 rue F. Joliot-Curie, 13388 Marseille Cedex 13  \\
$^{6}$ Department of Physics \& Astronomy, University of California, Irvine, CA 92697, USA \\
$^{7}$ Division of Physics, Astronomy and Mathematics, California Institute of Technology, Pasadena, CA, 91125, USA \\
$^{8}$  Zentrum f{\"u}r Astronomie der Universit{\"a}t Heidelberg, Institut f{\"u}r Theoretische Astrophysik, Albert-Ueberle Str. 2, D-69120 Heidelberg, Germany \\
$^{9}$ School of Physics and Astronomy, Cardiff University, Queens Buildings, The Parade, Cardiff, CF24 3AA, UK \\
$^{10}$ CNRS, Institut d'Astrophysique Spatiale, UMR 8617, 91405 Orsay, France \\
$^{11}$ Universit{\'e} Paris Sud, Institut d$'$Astrophysique Spatiale, UMR 8617, 91405 Orsay, France \\
$^{12}$ Department of Physics \& Astronomy, University of Sussex, Brighton, BN1 9QH, UK \\
$^{13}$ Laboratoire AIM, CEA, Universit{\'e} Paris VII, IRFU/Service d$'$Astrophysique, Bat. 709, 91191 Gif-sur-Yvette, France \\
$^{14}$ NASA Herschel Science Center, MS 100-22, California Institute of Technology, Pasadena, CA 91125, USA \\
$^{15}$ Istituto di Astrofisica e Planetologia Spaziali, INAF-IAPS, Via Fosso del Cavaliere 100, I-00133 Roma, Italy \\
$^{16}$ Department of Physics \& Astronomy, McMaster University, Hamilton, Ontario, L8S 4M1, Canada} 

   \date{Received ; accepted }

  \abstract
   {Dust reprocesses about half of the stellar radiation in galaxies. The thermal re-emission by dust of absorbed energy is considered driven merely by young stars and, consequently, often applied to trace the star formation rate in galaxies. Recent studies have argued that the old stellar population might anticipate a non-negligible fraction of the radiative dust heating.} 
   {In this work, we aim to analyze the contribution of young ($\lesssim$ 100 Myr) and old ($\sim$ 10 Gyr) stellar populations to radiative dust heating processes in the nearby grand-design spiral galaxy M\,51 using radiative transfer modeling. High-resolution 3D radiative transfer (RT) models are required to describe the complex morphologies of asymmetric spiral arms and clumpy star-forming regions and model the propagation of light through a dusty medium. } 
   {In this paper, we present a new technique developed to model the radiative transfer effects in nearby face-on galaxies. We construct a high-resolution 3D radiative transfer model with the Monte-Carlo code SKIRT accounting for the absorption, scattering and non-local thermal equilibrium (NLTE) emission of dust in M\,51. The 3D distribution of stars is derived from the 2D morphology observed in the IRAC\,3.6\,$\mu$m, \textit{GALEX} $FUV$, H$\alpha$ and MIPS\,24\,$\mu$m wavebands, assuming an exponential vertical distribution with an appropriate scale height. The dust geometry is constrained through the far-ultraviolet ($FUV$) attenuation, which is derived from the observed total-infrared-to-far-ultraviolet luminosity ratio. The stellar luminosity, star formation rate and dust mass have been scaled to reproduce the observed stellar spectral energy distribution (SED), $FUV$ attenuation and infrared SED.}
 {The dust emission derived from RT calculations is consistent with far-infrared and sub-millimeter observations of M\,51, implying that the absorbed stellar energy is balanced by the thermal re-emission of dust. The young stars provide 63$\%$ of the energy for heating the dust responsible for the total infrared emission (8-1000\,$\mu$m) while 37$\%$ of the dust emission is governed through heating by the evolved stellar population. In individual wavebands, the contribution from young stars to the dust heating dominates at \textit{ALL} infrared wavebands but gradually decreases towards longer infrared and submillimeter wavebands for which the old stellar population becomes a non-negligible source of heating. Upon extrapolation of the results for M\,51, we present prescriptions to estimate the contribution of young stars to the global dust heating based on a tight correlation between the dust heating fraction and specific star formation rate.} 
  {}

   \keywords{radiative transfer --
                dust, extinction --
                galaxies: individual: M\,51
                galaxies: ISM --
               infrared: galaxies}

   \maketitle
%

\section{Introduction}
Although dust only makes up a small part of the interstellar material (typically 1$\%$ in mass), its impact on the other interstellar medium (ISM) constituents becomes important in several astrophysical processes and galaxy evolution. Dust particles act as a catalyst in the formation process of molecular hydrogen on the surfaces of dust particles \citep{1971ApJ...163..155H}. The smallest dust grains regulate the heating of the neutral ISM gas component through photoelectric heating \citep{1994ApJ...427..822B} and inelastic interactions with gas particles. The irregular shape of large dust grains supplies shielding for molecules and ions from the hard radiation of young stars, hereby promoting the congregation and recombination of several elements, molecules and ions. 

Dust particles also distort our view on the other galaxy components (stars, gas, ...) through absorption and scattering processes. The location, distribution and structure of dust clouds will control the propagation of light. An accurate description of the obscuration effects of dust is, therefore, necessary to determine the specific amount of processed light by dust grains which will be re-emitted at infrared(IR)/submillimeter(submm) wavelengths. Only a proper characterization of the effects of dust across the multi-wavelength spectrum will allow us to recover and interpret several fundamental properties such as the star formation history (SFH), star formation rate (SFR), initial mass function (IMF), etc (e.g. \citealt{2007MNRAS.379.1022D,2009ApJ...691..394M,2010MNRAS.403.1894D}).

Other than corrections for dust obscuration to ultraviolet and optical data, it is also important to understand the different sources that heat the dust at infrared and submillimeter wavelengths. Especially because the far- and total-infrared emission in galaxies is often fully attributed to star formation and, thus, used to derive an estimate of the star formation activity (e.g. \citealt{2009ApJ...703.1672K}). While the radiative cooling through thermal emission at infrared wavelengths is the dominant cooling mechanism for dust, the heating of dust can be powered by different mechanisms (see also \citealt{1994ApJ...429..153S,2002AJ....124.3135K,2011AJ....142..111B,2012MNRAS.419.1833B}). Collisional heating, which dominates in hot gas where the electron velocities are high, and chemical heating, due to the formation of molecular hydrogen on the grain surface, are generally negligible on global galaxy scales compared to radiative heating mechanisms. 
Radiative heating can be driven by different stellar populations, ranging from young OB associations \citep{1990ApJ...350L..25D,1997AJ....113..599D} over evolved stars with mass losing photospheres or circumstellar envelopes \citep{1992ApJ...399...76K,1994ApJ...426...97M} to the general stellar radiation field. The latter emission spectrum is composed of energy escaping from active star-forming regions \citep{1990A&A...237..296S, 2000A&A...362..138P, 2001A&A...372..775M} as well as the underlying old stellar population \citep{1986ApJ...311L..33H,1994ApJ...429..153S,1996ApJ...456..163X,2002ApJ...572..232L,2004A&A...428..409B}.  
The dust grains heated by the old stars and light escaping from star-forming regions shows up as large-scale diffuse dust emission, often denoted as cirrus. In high-energy environments, the dust might be heated mechanically through shocks and/or turbulence in the ISM \citep{2006ApJ...651.1272M}, but its contribution is negligible on global galaxy scales. Other dust heating processes are powered by an active galactic nucleus (AGN) heating the surrounding dusty medium (e.g. \citealt{1985Natur.314..240D,2007ApJ...668...87W}) or cosmic rays (e.g. \citealt{1995ApJ...443..152W}). 

In the past, other works have employed surface brightness ratios of several infrared bands to study the main dust heating mechanisms by looking at their correlation with several star formation rate indicators and total stellar mass tracers (\citealt{2010A&A...518L..65B,2010A&A...518L..55G,2011AJ....142..111B,2012MNRAS.419.1833B,2014A&A...565A...4H,Bendo}). The study of a single infrared surface brightness ratio has the advantage of being mainly sensitive to dust temperature changes. The exploration of multi-waveband dust SED fitting techniques, however, benefits from the broader wavelength coverage to study the different heating sources in several infrared wavebands (e.g. \citealt{2009A&A...507..283M,2012ApJ...756..138A,2012ApJ...745...95D,2012ApJ...755..165M,2012ApJ...756...40S,2014A&A...565A.128C,2014arXiv1403.4272V}). Panchromatic SED fitting methods are even capable of constraining the absorbed energy from star light and re-emitted photons from dust grains by postulating a dust energy balance (e.g. CIGALE, MAGPHYS and GRASIL among others). 

The disadvantage of SED fitting is often the lack of stringent constraints on the 3D geometry of stars and dust in galaxies and the ignorance on the propagation distance and direction of light within a galaxy.
A complete self-consistent study of the dust heating mechanisms in galaxies requires high-resolution, 3D radiative transfer models (see \citealt{2013ARA&A..51...63S} for an overview on RT codes) accounting for the complex geometry of stars and dust in galaxies and directly solving the equation describing the transfer of stellar light through a dusty medium (e.g. \citealt{2000A&A...362..138P,2001A&A...372..775M,2011ApJ...738..124L}). Detailed RT models of spatially resolved galaxies, furthermore, provide the opportunity to analyze the spatial distribution of dust, the composition and emissivity of dust grains as well as possible variations throughout the galaxy. The main advantage of radiative transfer calculations is the non-local character of dust heating that can be addressed by tracing the propagation of stellar radiation through the dusty galaxy medium.

Up to the present day, RT modeling has been applied to a number of edge-on spiral galaxies \citep{2001A&A...372..775M,2004A&A...425..109A,2005A&A...437..447D,2008A&A...490..461B,2010A&A...518L..39B,2011A&A...527A.109P,2012MNRAS.419..895D,2012MNRAS.427.2797D,2012ApJ...746...70S,2013A&A...550A..74D,2014MNRAS.441..869D}, which have the advantage that the extinction and emission of dust can easily be observed along the line-of-sight and, that the dust can be vertically resolved and traced out to large radii. One drawback of modeling edge-on galaxies is the lack of insight in the spatial distribution of star-forming regions and the clumpiness of the interstellar medium, which impedes a spatially resolved analysis of the various stellar populations and dust grains and prevents analyzing the interplay between stars and dust to characterize the main dust heating mechanisms. To fully recover the asymmetric stellar and dust geometries, we require detailed radiative transfer calculations of well-resolved low-inclination systems, where the dust distribution can be studied into great detail and the presence of unobscured/embedded star-forming regions can be identified.
High-resolution 3D radiative transfer modeling of the Milky Way has allowed \citet{2012A&A...545A..39R} to identify a deficit in the dust distribution in the inner Galaxy and a potentially increased abundance of polycyclic aromatic hydrocarbons (PAHs). Owing to the improved spatial resolution and extended wavelength coverage down to the submillimeter regime provided by \textit{Herschel} \citep{2010A&A...518L...1P}, the radiative transfer modeling of stars and dust in nearby galaxies beyond our own Galaxy now becomes possible.

In this work, we aim to expand dust radiative transfer modeling of edge-on spirals galaxies to face-on counterparts. 
Due to the complex morphology with asymmetric spiral arms, bar and ring structures, obscured star-formation regions and the clumpy nature of the interstellar medium visible in face-on disk galaxies, we require a high-resolution 3D radiative transfer model that can simultaneously reproduce the multi-wavelength emission of stars and dust across the electromagnetic spectrum.
The complexity of 3D radiative transfer models with their high number of unknown parameters in combination with the computational cost of the full radiative transfer treatment makes it a challenging task. Given that the asymmetric structures of face-on galaxies are hard to parameterize with analytic functions and that the clumpy nature of localized sources within the galaxy's disk will be a hard nut to crack for most standard optimization techniques (e.g. genetic algorithms, Levenberg-Marquardt minimization, etc.), we present an alternative modeling technique that will benefit from high-resolution imaging observations to constrain the stellar and dust geometries in the radiative transfer model.

In this paper, we construct a 3D radiative transfer model of the grand-design spiral galaxy M\,51\footnote{M\,51 is a nearby, late-type (SAbc) galaxy located at a distance of $D$ = 8.4 Mpc \citep{1997ApJ...479..231F}, and belongs to an interacting system with the companion galaxy NGC\,5195. The galaxy pair is often referred to as M\,51, while the individual galaxy NGC\,5194 is named M\,51a. In this paper, we refer to M\,51 for the single galaxy NGC\,5194 belonging to the interacting system.} (NGC\,5194) with a near to face-on orientation ($i$ $\sim$ 20$^{\circ}$) and a huge database of multi-band ancillary data. We exploit the panchromatic RT model of stars and dust in M\,51 to analyze the main dust heating sources on spatially resolved scales across the infrared spectrum. Rather than giving hints on the predominant dust heating source, the RT calculations track the emission of stellar populations of different ages and their interaction with the dusty interstellar medium, which allows to constrain the percentage of dust heated by young stars ($\lesssim$ 100 Myr) and the old stellar population ($\sim$ 10 Gyr) at every location in 3D space across multiple infrared wavebands. 

The paper is organized with the description of the multi-wavelength dataset presented in Section \ref{Data.sec}. In Section \ref{RTmodel}, we introduce a new method for radiative transfer modeling to constrain the geometry of stars and dust in face-on galaxies based on observations in ultra-violet (UV), optical, near-infrared (NIR) and infrared (IR) wavebands. We discuss the strengths and possible flaws of the constructed RT model in Section \ref{Fitting.sec}. Section \ref{Analysis.sec} applies the RT model to analyze the main dust heating mechanisms in M\,51 and the variation in dust heating sources with infrared wavebands. The main conclusions of the paper are summed up in Section \ref{Con.sec}. In appendices \ref{effect_scaleheight} and \ref{effect_clumps}, we investigate the implications introduced by the model assumptions on the relative dust-to-stellar scale height and clumpiness of the dust distribution, respectively, for the RT model of M\,51.

\section{Multi-wavelength dataset}
\label{Data.sec}
The paper aims to construct a multi-wavelength radiative transfer model for M\,51 that can self-consistently explain the stellar emission, dust attenuation and, dust emission. We, hereto, present a new RT modeling technique that employs observations to characterize the geometrical distribution of stars and dust in a 3D model. To compare the model output images with observations, we require a multi-wavelength dataset. The observational dataset used for model construction (see Section \ref{Model1.sec}) and model validation (see Section \ref{Fitting.sec}) is described in Section \ref{stellar.sec}, while the necessary data pre-processing steps are outlined in Section \ref{process.sec}. 

\subsection{Dataset}
\label{stellar.sec}

Ultra-violet (UV) maps were obtained with the Galaxy Evolution Explorer (\textit{GALEX}, \citealt{2005ApJ...619L...1M}) with a total observing time of 1414 s as part of the \textit{GALEX} Nearby Galaxy Survey (NGS, \citealt{2003lgal.conf...10B,2003AAS...203.9112B,2004AAS...205.4201G}), a program collecting deep \textit{GALEX} $FUV$ and $NUV$ data for about 200 nearby galaxies. We collected \textit{GALEX} $FUV$ and $NUV$ maps provided by \citet{2007ApJS..173..185G} from the NASA Extragalactic Database (NED)\footnote{http://ned.ipac.caltech.edu/}. 

H$\alpha$ data were retrieved from \citet{2002A&A...386..124B}. The galaxy was observed with the 1.20 m Newton telescope at the Observatoire de Haute Provence using a TK $1024\times1024$ pixel CCD detector and narrowband filters at 6450 and 6561$\AA$ with exposure times of 900s each. After standard data processing steps were applied to the data, the 6450 $\AA$ image was subtracted from the 6561$\AA$ image to produce the H$\alpha$ image. Foreground stars were identified as negative point-like residual features and were removed by interpolating over the data. The final image has pixels with sizes of 0.69$\arcsec$, a calibration uncertainty of 5$\%$, and a PSF FWHM of 2$\arcsec$.

M\,51 was also observed as part of the Sloan Digital Sky Survey (SDSS) in two separate frames (each with an exposure time of 54 s) using the $u$, $g$, $r$, $i$ and $z$ filters. We downloaded the two image tiles from the SDSS DR7 archive \citep{2009ApJS..182..543A}. The two separate image frames are background subtracted before they are combined to final mosaics based on standard \texttt{iraf} image processing tasks (\texttt{imalign}, \texttt{imcombine}).

The Two Micron All Sky Survey (2MASS, \citealt{2003AJ....125..525J}) observed M\,51 in the $J$, $H$, $K_{s}$ filters as part of the Large Galaxy Atlas Survey. We collected 2MASS $J$, $H$, $K_{s}$ maps provided by \citet{2003AJ....125..525J} from NED. 

The Wide-field Infrared Survey Explorer (WISE, \citealt{2010AJ....140.1868W}) observed M\,51 in its four photometric bands at 3.4, 4.6, 11.6 and 22.0\,$\mu$m. WISE maps are converted from DN units to Jy based on the photometric zeropoints quoted in the User's Guide to the WISE Preliminary Data Release\footnote{http://wise2.ipac.caltech.edu/docs/release/prelim/expsup/sec2\_3f.html}. Three types of corrections are recommended for extended source photometry by \citet{2013AJ....145....6J}. The first aperture correction factor corrects for the PSF profile fitting that is used for the WISE absolute photometric calibration, with corrections of 0.034, 0.041, -0.030 and 0.029 mag applied to the WISE 3.4, 4.6, 11.6 and 22.0\,$\mu$m filters. The second correction corrects for the shape of the spectrum. We do not account for this colour correction since the SKIRT output spectra are convolved with the appropriate WISE response filters. The last correction factor only applies to the WISE 4 band (0.92), correcting for the calibration discrepancy between WISE photometric standard blue stars and red galaxies.

The Infrared Array Camera (IRAC, \citealt{2004ApJS..154...10F}) and Multiband Imaging Photometer (MIPS, \citealt{2004ApJS..154...25R}) on board the \textit{Spitzer} Space telescope observed M\,51 as part of the \textit{Spitzer} Infrared Nearby Galaxy Survey (SINGS, \citealt{2003PASP..115..928K}). We retrieved IRAC\,3.6, 4.5, 5.8 and 8.0\,$\mu$m maps from the SINGS data archive\footnote{http://irsa.ipac.caltech.edu/data/SPITZER/SINGS/}. We, furthermore, retrieve an optical $R$ band image from the SINGS data archive. The IRAC images were multiplied with the appropriate extended source photometrical correction coefficients (0.91, 0.94, 0.66 and 0.74 for the IRAC\,3.6, 4.5, 5.8 and 8\,$\mu$m wavebands; see the IRAC instrument cookbook\footnote{http://irsa.ipac.caltech.edu/data/SPITZER/docs/irac/iracinstrument\-handbook/IRAC\_Instrument\_Handbook.pdf}). 
We use the reprocessed MIPS 24\,$\mu$m data from \citet{2012MNRAS.423..197B}. 

M\,51 is observed by the \textit{Herschel} Space Observatory \citep{2010A&A...518L...1P} as part of the Very Nearby Galaxy Survey (VNGS, PI: Christine Wilson). The PACS (70, 160\,$\mu$m, ObsID 1342188328, 1342188329) and SPIRE (250, 350, 500\,$\mu$m, ObsID 1342188589) instruments onboard \textit{Herschel} performed continuum mapping of the interacting galaxy pair NGC\,5194/NGC\,5195. 
PACS and SPIRE data of M\,51 were observed on December 20th and 26th 2010, covering an area of about 20$\arcmin \times 20\arcmin$ centered on the galaxy M\,51.
The same area was observed in nominal and orthogonal scan direction with four and two repetitions with PACS and SPIRE instruments, respectively, at a medium scan speed (20 arcsec s$^{-1}$).
The \textit{Herschel} dataset was already presented in \citet{2012ApJ...755..165M}. For the analysis in this paper, PACS data were retrieved from the \textit{Herschel} Science Archive (HSA) and re-processed using the \textit{Herschel} Interactive Processing Environment (HIPE, v12, \citealt{2012ASPC..461..733B}) up to level 1 using the calibration file v56. The level 1 data are reduced in version 23 of \texttt{scanamorphos} \citep{2013PASP..125.1126R}. The SPIRE data reduction was performed in a similar way by processing the data up to level 1 using the standard pipeline and calibration file version \texttt{spire\_cal\_12\_1} in HIPE v11 and reducing the level 1 data using the BRIght Galaxy ADaptive Element method (Smith et al. in prep.). The latter uses a custom method to remove the temperature drift and bring all bolometers to the same level (instead of the default \textit{temperatureDriftCorrection} and the residual, median baseline subtraction). The SPIRE maps are multiplied with $K_{\text{PtoE}}$ correction factors to convert from point source to extended source photometric calibration. The appropriate correction factors (i.e. 91.289, 51.799, 24.039 at 250, 350 and 500\,$\mu$m, respectively; see SPIRE Observers' manual) for a constant $\nu S_{\nu}$ spectrum convert the maps from Jy/beam to MJy/sr units, accounting for the most up-to-date measured SPIRE beam areas of 465, 823, 1769 arcsec$^2$ at 250, 350 and 500\,$\mu$m, respectively. 
No colour corrections are applied since the SKIRT output SED and images are convolved with the appropriate response curve to derive modeled band fluxes.

The full-width-half-maximum (FWHM) of the point-spread-function (PSF) in each of the different bands as well as the calibration uncertainties are summarized in Table \ref{band} along with the appropriate references reporting on these numbers.

\subsection{Data pre-processing}
\label{process.sec}

Foreground stars are identified from the 2MASS Point Source Catalog.
Similar to the correction techniques applied in \citet{2012ApJ...755..165M}, we only identify point sources as stars (and not H{\sc{ii}} regions) if the surface brightness ratio of the IRAC\,8\,$\mu$m band versus $R$ band emission, $I_{\text{8}}$/$I_{\text{R}}$, is lower than 1.5.
Point sources satisfying this criterion are removed, i.e. their emission is replaced by the median pixel value in the background annulus with inner radius 8$\arcsec$ and outer radius 16$\arcsec$ around the star's position.

The background in each of the maps has been determined by measuring the background in a sufficiently large number (depending on the size of the background region available) of apertures of size $R=4$ $\times$ FWHM, randomly distributed outside the bright emission regions of M\,51. The final background value is chosen as the median value of those measurements, with the standard deviation being representative for the uncertainty on the background subtraction procedure. The background value has been subtracted from the frames in case the median background value exceeded the standard deviation.

All maps are corrected for Galactic extinction according to the re-calibrated $A_{\lambda}$ in \citet{2011ApJ...737..103S} from \citet{1998ApJ...500..525S}, as reported on NED. The attenuation in the ultraviolet wavebands is determined based on the $V$-band attenuation $A_{\text{V}}$ and assuming an extinction law with $R_{\text{V}}$ derived in \citet{1999PASP..111...63F}.

To homogenize the observational dataset and reconcile the resolution of the input geometries in SKIRT, we convolve the entire dataset to the resolution of the PACS\,160\,$\mu$m image ($\sim$ 12.1$\arcsec$ or 493 pc at the distance of M\,51), which is the lowest resolution map that will be used to constrain stellar and dust geometries in SKIRT\footnote{We prefer to convolve all data to the resolution of the PACS\,160\,$\mu$m map to maintain the spatial structures at scales of $\sim$ 500 pc rather than convolving all data to the resolution in the SPIRE wavebands. For  comparison between SPIRE modeled and observed data, we convolve the model images to the appropriate SPIRE resolution.}. The appropriate convolution kernels are retrieved from \citet{2011PASP..123.1218A}. All convolved images are rebinned to maps with pixel size of 2$\arcsec$. With a pixel size six times smaller compared to the FWHM of the PSF in the PACS\,160\,$\mu$m waveband ($\sim$ 12.1$\arcsec$), adjacent pixels will not be independent in the input images describing the stellar and dust geometries in SKIRT. Given that the non-local interactions between stellar photons and dust particles are inherent to radiative transfer calculations, the possible dependence of adjacent pixels in the input geometries will not severely affect the RT models. 

To make the size of each input image consistent, we restrict the emission in the PACS\,160\,$\mu$m to pixels detected at a signal-to-noise level of at least five, resulting in 35,375 pixels of size 2$\times$2 arcsec$^{2}$. The emission in the other input images is cut at the same contour level to prevent that the extent of the SKIRT input images and, thus, the radial distribution differs among RT components.

\begin{table}
\caption{Overview of the characteristics (full-width half maximum of the beam, calibration uncertainties) for the multi-wavelength dataset used in the RT analysis of M\,51.}
\label{band}
\centering
\begin{tabular}{lccccc}
\hline 
\hline 
Band & $\lambda$ & FWHM & Ref\tablefootmark{a}  & $\sigma_{cal}$ & Ref\tablefootmark{a} \\ 
& [$\mu$m] & [arcsec] & & & \\
\hline 
\textit{GALEX} $FUV$ & 0.15 & 4.2 & 1 & 0.05 mag & 1  \\
\textit{GALEX} $NUV$ & 0.23 & 5.3 & 1 & 0.03 mag & 1  \\
SDSS $u$ & 0.35 & 2\tablefootmark{b} & 2 & 2$\%$ & 3  \\
SDSS $g$ & 0.48 & 2\tablefootmark{b} & 2 & 2$\%$ & 3  \\
SDSS $r$ & 0.62 & 2\tablefootmark{b} & 2 & 2$\%$ & 3   \\
H$\alpha$ & 0.66 & 2 & 4 & 5$\%$ & 4 \\
SDSS $i$ & 0.76 & 2\tablefootmark{b} & 2 & 2$\%$ & 3   \\
SDSS $z$ & 0.91 &  2\tablefootmark{b} & 2 & 2$\%$ & 3  \\
2MASS $J$ & 1.25 & 3.3\tablefootmark{c}  & 2 & 0.03 mag & 5 \\
2MASS $H$ & 1.65 & 3.1\tablefootmark{c}  & 2 & 0.03 mag & 5 \\
2MASS $K$ & 2.16 & 3.3\tablefootmark{c}  & 2 & 0.03 mag & 5   \\
\textit{WISE}\,1 & 3.4 & 6.1 & 6 & 2.4$\%$ & 7 \\ 
IRAC\,3.6\,$\mu$m & 3.6 & 1.7 & 8 & 1.8$\%$ & 9 \\
IRAC\,4.5\,$\mu$m & 4.5 & 1.7 & 8 & 1.9$\%$ & 9 \\
\textit{WISE}\,2 & 4.6 & 6.4 & 6 & 2.8$\%$ & 7 \\ 
IRAC\,5.8\,$\mu$m & 5.8 & 1.9 & 8 &  2.0$\%$ & 9 \\
IRAC\,8.0\,$\mu$m & 8.0 & 2.0 & 8 &  2.1$\%$ & 9 \\
\textit{WISE}\,3 & 12 & 6.5 & 6 & 4.5$\%$ & 7 \\ 
\textit{WISE}\,4 & 22 & 12 & 6 & 5.7$\%$ & 7 \\ 
MIPS\,24 & 24 & 6 & 10 & 4$\%$ & 10 \\
PACS\,70 & 70 & 5.8 & 11 & $5\%$ & 12 \\
PACS\,160 & 160 & 12.1 & 11 & $5\%$ & 12 \\
SPIRE\,250 & 250 & 18.2 & 13 & $4\%$\tablefootmark{d} & 13 \\
SPIRE\,350 & 350 & 24.9 & 13 & $4\%$\tablefootmark{d} & 13 \\
SPIRE\,500 & 500 & 36.3 & 13 & $4\%$\tablefootmark{d} & 13 \\
\hline 
\end{tabular}
\tablefoot{
\tablefoottext{a}{References: (1) \citet{2007ApJS..173..682M}; (2) this paper; (3) \citet{2008ApJ...674.1217P};  (4) \citet{2002A&A...386..124B}; (5) \citet{2006AJ....131.1163S}; (6) \citet{2010AJ....140.1868W}; (7) \citet{2011ApJ...735..112J}; (8) \citet{2004ApJS..154...10F}; (9) \citet{2005PASP..117..978R}; (10) \citet{2007PASP..119..994E}; (11) PACS Observers' Manual; (12) \citet{2013ExA...tmp...38B} ;(13) SPIRE Observers' Manual. }\\
\tablefootmark{b}{We estimate the FWHM of the SDSS data by fitting a 2D Gaussian (IDL task \texttt{starfit}) to several stellar objects in the field of M\,51. We apply the sum of two Gaussian kernels with FWHM of 2$\arcsec$ to convolve the SDSS data, which should be a good representation of the telescope's PSF \citep{2011PASP..123.1218A}. }\\
\tablefootmark{c}{We estimate the FWHM of the 2MASS data by fitting a 2D Gaussian (IDL task \texttt{starfit}) to several stellar objects in the field of M\,51. We apply Gaussian kernels with FWHM of 3$\arcsec$ ($H$) and 3.5$\arcsec$ ($J,K$) to convolve the 2MASS data to the PACS\,160\,$\mu$m resolution \citep{2011PASP..123.1218A}.}\\
\tablefootmark{d}{Calibration uncertainties for the \textit{Herschel} SPIRE instrument are assumed to be around 4$\%$ in each band, adding in quadrature the 4$\%$ absolute calibration error from the assumed models used for Neptune (SPIRE Observers' manual) and a random uncertainty of 1.5$\%$ accounting for the repetitive measurements of Neptune (see \citealt{2013MNRAS.433.3062B}).}}
\end{table}

\section{3D radiative transfer model}
\label{RTmodel}

\subsection{Radiative transfer code: SKIRT}
\label{SKIRT.sec}
SKIRT \citep{2003MNRAS.343.1081B,2011ApJS..196...22B} is a 3D Monte Carlo radiative transfer code designed to model the absorption, scattering and thermal
re-emission of dust in a variety of environments: circumstellar disks \citep{2007BaltA..16..101V}, clumpy tori around active galactic nuclei \citep{2012MNRAS.420.2756S} and different galaxy types \citep{2010A&A...518L..39B,2010MNRAS.403.2053G,2010A&A...518L..45G,2012MNRAS.419..895D,2012MNRAS.427.2797D,2013A&A...550A..74D}. SKIRT can furthermore be applied to model the emission of simulated galaxies from hydrodynamical simulations \citep{2013ASPC..474..53M}.
An important extension of SKIRT in view of high-resolution radiative transfer models involves the implementation of adaptive grid structures such as the hierarchical Octree \citep{2001A&A...379..336K,2006A&A...456....1N}, $k$-d grid structures and Voronoi tesselations \citep{Voronoi} which are described in more detail in \citet{2013A&A...554A..10S}, \citet{2014A&A...561A..77S} and \citet{2013A&A...560A..35C}, respectively.
Such adaptive grid structures allow to increase the resolution in the dust grid where dust clumps or asymmetric spiral arm structures are present, and at the same time keep the computational cost limited.

\subsection{Model construction}
\label{Model1.sec}

The construction of a 3D radiative transfer model composed of stars and dust includes the characterization of the \textit{geometrical distribution}, \textit{spectrum} and \textit{normalization} of each of the different stellar and dust components.
The geometrical distribution of SKIRT components is described through 3D geometries that render the probability density distribution for stars or dust (see Section \ref{Geometry.sec}). These 3D density distributions give the probability that a given location in 3D space will be populated by dust particles or stars of a certain age.
The description of the emission spectra of stars characterizes the variation of stellar emission with wavelength (see Section \ref{Spectra.sec}), while the normalization of the spectra encompasses the scaling of this emission spectrum for a given bolometric luminosity or a luminosity specified in a given waveband (see Section \ref{Normalization.sec}). The emission spectrum for the dust component in our RT calculations is constrained  through the grain composition, with the grain size distribution, optical dust properties and dust emissivities specified for each individual grain population (see Section \ref{Spectra.sec}), and normalized by the total dust mass (see Section \ref{Normalization.sec}).

\subsubsection{SKIRT model distribution: 3D stellar and dust geometries}
\label{Geometry.sec}
\begin{figure}
\centering
\includegraphics[width=8.5cm]{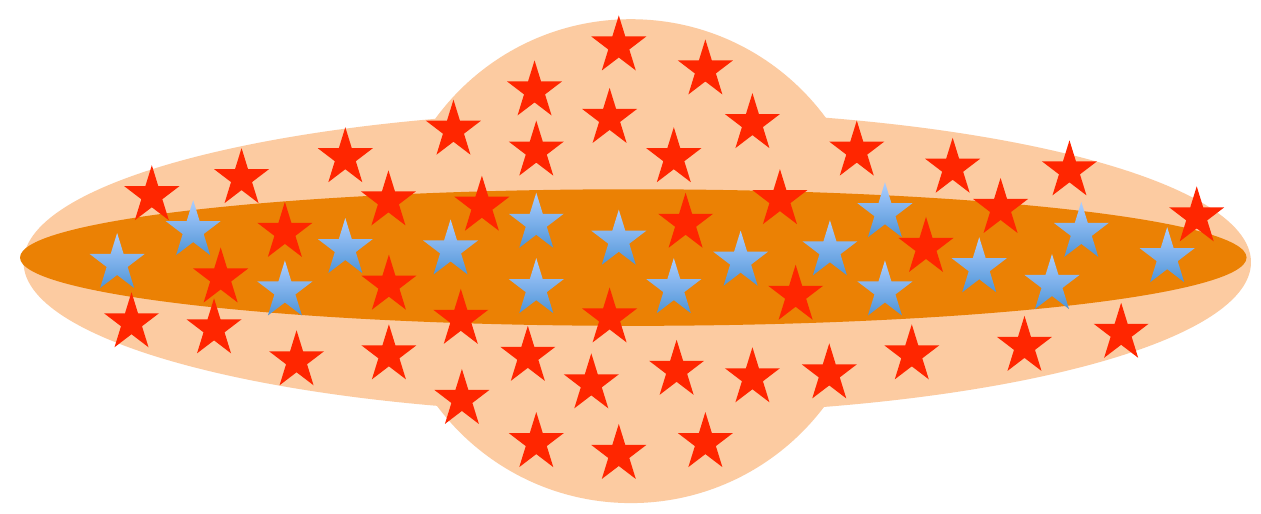}   \\
 \caption{A schematic diagram of our standard model observed from an edge-on view, consisting of a bulge and thick disk component (represented with a light orange color) of old stars (red stars) and a thin, star-forming disk (dark orange) containing a dust component and young stars (blue stars).} 
               \label{plot_general_model}
\end{figure}

The model construction for M\,51 is based on the standard model represented in Figure \ref{plot_general_model}.
The standard model is composed of a bulge, thick and thin disk with the thin disk harboring the dust and young stars while the bulge and thick disk consist merely  of old stars. The standard model represented in Figure \ref{plot_general_model} has been applied -whether or not with some small modifications- in several other radiative transfer studies  (e.g. \citealt{1999A&A...344..868X,2000A&A...362..138P,2004ApJ...617.1022P,2004A&A...419..821T,2008A&A...490..461B,2011A&A...527A.109P}) and has been proven to give a realistic representation of the geometry of stars and dust in galaxies, capable of reproducing the observed UV/optical and IR/submm emission. While the standard model also applies to model edge-on galaxies, the galaxies observed from a face-on perspective require 3D asymmetric geometries to describe the observed stellar and dust structures, which is different from the 2D analytic prescriptions used to describe the stars and dust in edge-on galaxies. 
 
The geometrical distribution of the stellar and dust components in the 3D radiative transfer model of M\,51 is constrained based on a set of multi-wavelength observations. Although observations only depict a 2D projection of the 3D galaxy, we are able to recover the 3D geometry through de-projection following an Euler-like transformation described by the equation:
\begin{equation}
\begin{array}{lcl}
\begin{pmatrix} x_{p} \\ y_{p} \\ z_{p} \end{pmatrix} & = & \begin{pmatrix} \cos~\omega & -\sin~\omega & 0 \\  \sin~\omega & \cos~\omega & 0 \\ 0 & 0 & 1 \end{pmatrix} 
\begin{pmatrix} 0 & 1 & 0 \\  -1 & 0 & 0 \\ 0 & 0 & 1 \end{pmatrix} \times \\
& & \begin{pmatrix} \cos~\theta & 0 & -\sin~\theta \\ 0 & 1 & 0 \\  \sin~\theta & 0 & \cos~\theta  \end{pmatrix} 
\begin{pmatrix} \cos~\phi & -\sin-\phi & 0 \\  \sin-\phi & \cos~\phi & 0 \\ 0 & 0 & 1 \end{pmatrix}  \begin{pmatrix} x \\ y \\ z \end{pmatrix}
\end{array} 
\end{equation}
with an inclination angle $\theta$ $\sim$ 20$^{\circ}$, position angle $\omega$ $\sim$ 170$^{\circ}$ \citep{1974ApJS...27..437T}, azimuth $\phi$ $\sim$ 0$^{\circ}$, and the projected $z$ coordinate, $z_{\text{p}} = 0$. To describe the vertical $z$ distribution, we assume an exponential distribution $\exp(-|\text{z}|/\text{h}_{\text{z}})$ with a certain scale height, $\text{h}_{\text{z}}$.

Due to the nearly face-on view on M\,51 ($i=20^{\circ}$), we lack any constraints on the vertical scale height of stars and dust. Therefore, we need to rely on studies of the vertical extent of stars and dust in highly-inclined disk galaxies under the assumption that the relative star-dust geometry in M\,51 is similar to the observed vertical structures in high-inclinations systems. The scale heights have been shown to be comparable for stars and dust in low mass galaxies \citep{2005AJ....130.1574S}, but for high mass galaxies the ISM is considered to collapse into dust lanes distributed in thin disks \citep{2004ApJ...608..189D}. Also model fitting to several edge-on galaxies have resulted in smaller dust scale heights compared to the vertical extent of the thick disk composed of old stars \citep{1994MNRAS.270..373J,1999A&A...344..868X,2007A&A...471..765B,2013A&A...550A..74D,2013A&A...556A..54V,2014MNRAS.441..869D}. 
The scale height of dust in galaxies is, however, still debated with several galaxies showing evidence of extra-planar dust (e.g. \citealt{1999AJ....117.2077H,2000A&A...356..795A,2004AJ....128..662T}). While this extra-planar dust component is interpreted as infalling or outflowing material, the presence of small dust grains and polycyclic aromatic hydrocarbons (PAHs) traced up to several kpc above the plane of several edge-on spiral galaxies suggests that a diffuse dust component with scale heights exceeding the vertical scale height of more commonly observed stellar and dust disks might be present in galaxies and, thus, not of external origin \citep{2006A&A...445..123I,2007A&A...474..461I,2007A&A...471L...1K,2009MNRAS.395...97W}. The divergent angular resolution of observations at various infrared/submillimeter wavebands furthermore complicates the determination of dust scale heights. On top of the resolution issues, the intrinsic dust distribution is tough to define from the emission in a single waveband due to the different heating sources that might contribute at individual wavebands. 

Consistent with observational constraints and modeling results obtained for edge-on spiral galaxies, we assume that the old stars and dust are distributed in a disk with an exponential vertical profile with scale heights $h_{z,\star} \sim 450$ pc and $h_{\text{z},d} \sim 225$ pc, respectively. The scale height of old stars in M\,51 ($h_{z,\star}$ $\sim$ 450 pc) is based on the scale length fitted to the observed $R$, $I$ and $K$ band images ($h_{\text{R},\star} \sim$ 90$\arcsec$ or 3665 pc, \citealt{1996ApJ...467..175B}) and the mean intrinsic flattening of the stellar disk as obtained from 2D model decomposition ($h_{\text{R},\star}$/$h_{\text{z},\star}$ $\sim$ 8.21, \citealt{2002MNRAS.334..646K}) and 3D RT modeling ($h_{\text{R},\star}$/$h_{\text{z},\star}$ 8.26, \citealt{2014MNRAS.441..869D}). We assume a relative stellar-to-dust scale height ratio of 2:1, consistent with the RT model predictions for edge-on galaxies from \citet{1999A&A...344..868X,2007A&A...471..765B} and \citet{2014MNRAS.441..869D}, resulting in a dust scale height of 225 pc.  We study the effect of variations in the relative dust-to-stellar scale heights in Section \ref{effect_scaleheight}, which shows that changing the relative dust-to-stellar scale height results in an inconsistency with the observed $FUV$ attenuation and IR emission. We are, therefore, confident that the assumed relative dust-to-stellar scale height $h_{\text{z,d}}$/$h_{\text{z,}\star}$ = 0.5 provides a good representation of the vertical distribution of stars and dust in M\,51. The comparison of 3D radiative transfer models with variable dust-to-stellar scale heights suggests that the observed $FUV$ attenuation is able to put stringent constraints on the relative dust-to-stellar vertical distribution in galaxies. This new technique of determining the relative dust-to-stellar scale height will be exploited for a larger sample of nearby galaxies in future work.

The young stars (both ionizing and non-ionizing) are assumed to be distributed in a thin disk with scale height $h_{\text{z}} \sim 100$ pc, which is compatible with the scale height of star-forming disks in our Galaxy \citep{1980ApJS...44...73B} and NGC\,891 \citep{2013ApJ...773...45S}. A power spectrum analysis of the dust emission at 24 and 100\,$\mu$m in M\,33, furthermore, results in similar dust scale heights of 100\,pc for dust components heated by young stars \citep{2012A&A...539A..67C}.

The scale heights of stellar and dust disks are assumed to be invariant throughout the disk of M\,51. The stellar scale height has been shown to increase with galactocentric distance in some galaxy disks (e.g. \citealt{1997A&A...320L..21D,2002A&A...390L..35N}), but this variation should mainly occur in early-type galaxies and become negligible in late-type spirals \citep{1997A&A...320L..21D}. The recent interaction of NGC\,5194 with companion NGC\,5195 \citep{2000MNRAS.319..377S,2010MNRAS.403..625D} might, however, have caused twisting or warping of the disk (e.g. \citealt{2007ApJ...665.1138S}), resulting in a changing position angle and/or inclination with radius.

Hereafter, we specify the observational constraints used to describe the geometrical distribution of the different stellar and dust components in the standard model (see also Table \ref{SKIRTcomp}). 

\textbf{Old stellar population} \\
The old stellar component is constrained by the observed IRAC\,3.6\,$\mu$m emission. However, we need to separate the emission originating from the bulge and disk components in this IRAC\,3.6\,$\mu$m map. Hereto, we model the stellar bulge using our radiative transfer code SKIRT and assuming the best fitting bulge parameters reported by \citet{2010AJ....139.2097P}, based on their 2D model decomposition of M\,51 derived with the galaxy fitting algorithm \texttt{galfit} \citep{2002AJ....124..266P}. More specifically, the emission of the bulge is modeled using a flattened Sersic profile with a Sersic index $n$ = 0.67, effective radius $R_{\text{e}}$ = 635.3 pc and flattening parameter $q$ = 0.88, assuming an inclination angle of 20$^{\circ}$ and position angle of 170$^{\circ}$. We scale the luminosity of the bulge until we reproduce the bulge-to-disk ratio (0.16) of the model for M\,51 presented by \citet{2010AJ....139.2097P}. After subtraction of the bulge component, the IRAC\,3.6\,$\mu$m map is considered to only include the emission of old stars distributed in the thick disk.

\textbf{Young non-ionizing stars} \\
We rely on the observed \textit{GALEX} $FUV$ image, dominated by emission from B and more massive A stars tracing the unobscured star formation over time scales of 10-100 Myr \citep{1999ApJ...521...64M,2009ApJ...706.1527B}, to constrain the geometrical distribution of the young non-ionizing stellar component in M\,51. 

\textbf{Young ionizing stars} \\
We rely on the observed H$\alpha$ and MIPS\,24\,$\mu$m maps to describe the 2D distribution of young ionizing stars. Given that part of the 24\,$\mu$m emission might result from diffuse dust heated by the interstellar radiation field rather than star-forming regions, we need to correct for this diffuse dust component contributing to the 24\,$\mu$m emission. We estimate that the global emission of diffuse regions contributing to the 24\,$\mu$m emission amounts to about 20$\%$ - assuming that the contribution is similar to the diffuse dust component observed in the Scd galaxy M\,33 (see \citealt{2012A&A...543A..74X})\footnote{Based on the contribution of old stars to the 24\,$\mu$m (35$\%$) emission derived for the Sab galaxy M\,81 \citep{Lu}, the diffuse emission component emitting at 24\,$\mu$m might become higher in spiral galaxies with an earlier-type classification and, thus, possibly be also higher in M\,51 compared to the 20$\%$ derived for M\,33.}.
We model an exponential disk with scale length $h_{\text{R}} = 2000$ pc (following \citealt{2012A&A...543A..74X}) with the radiative transfer code SKIRT, scale it to 20$\%$ of the global 24\,$\mu$m flux of M\,51 and subtract the diffuse disk component from the observed 24\,$\mu$m map. A map of young ionizing stars is constructed based on the prescriptions from \citet{2007ApJ...666..870C}\footnote{We prefer to rely on the calibration coefficient of 0.031 reported by \citet{2007ApJ...666..870C} in combination with the diffuse emission-subtracted MIPS\,24\,$\mu$m map rather than the factor 0.02 given by \citet{2009ApJ...703.1672K} which accounts for the contribution from diffuse emission heated by old stellar populations.} to correct the H$\alpha$ emission for internal extinction, i.e. $L_{H\alpha,\text{corr}}$ = $L_{H\alpha}$ + 0.031 $\times$ $L_{24}$.

\begin{figure*}
\centering
\includegraphics[width=18.5cm]{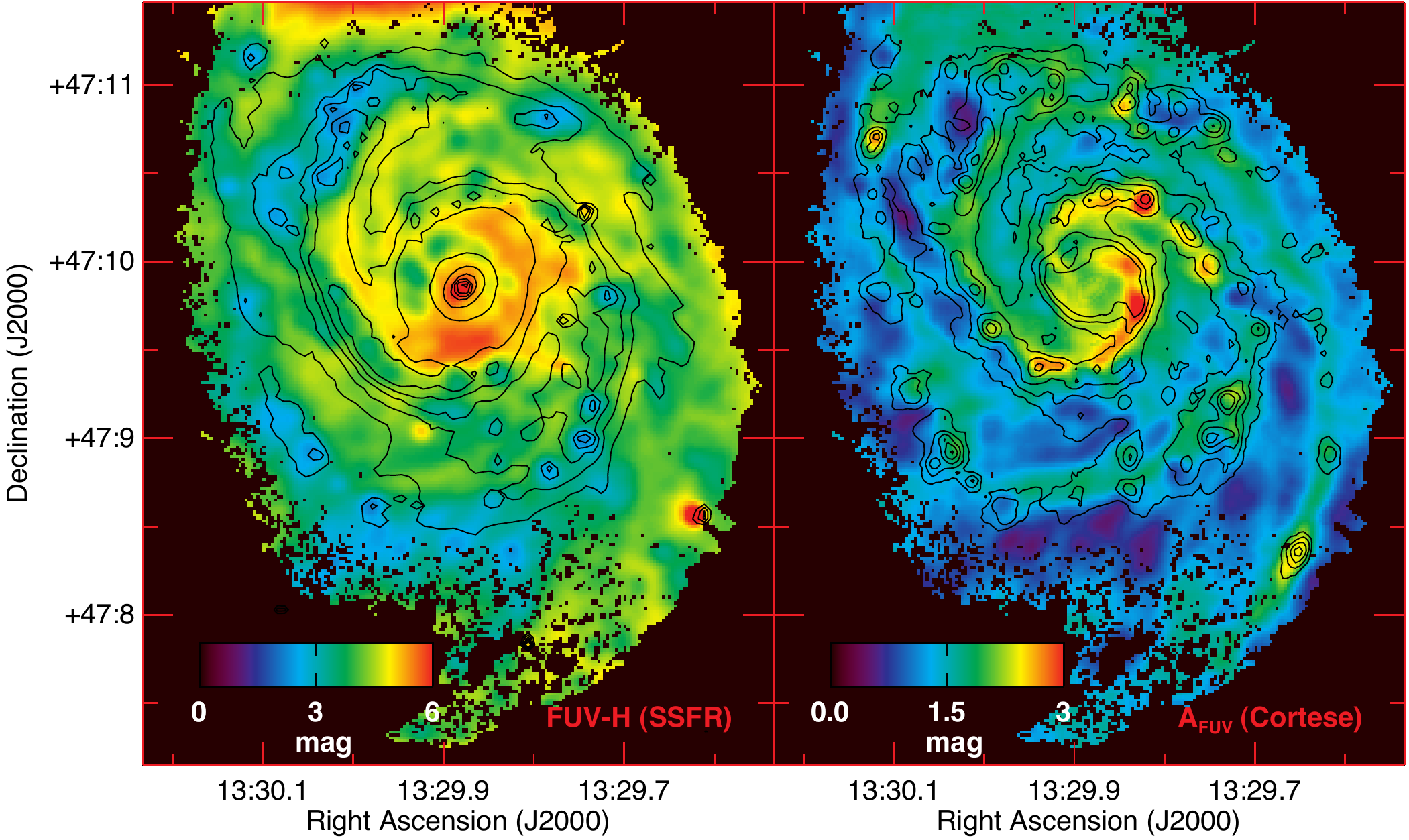}   \\
 \caption{The $FUV-H$ color map (left), which is considered a proxy of the specific star formation rate, and the $FUV$ dust attenuation (right) in M\,51. The $R$ band contours has been overlaid as black contours in the left panel, while the black contours in the right panel trace the dust distribution as observed from the PACS\,70\,$\mu$m image.}
              \label{plot_Cortese}
\end{figure*}

\textbf{Dust component} \\
\label{Dustcomp.sec}
The geometry of the dust component in M\,51 is constrained through the $FUV$ attenuation. The $FUV$ attenuation is computed based on the calibrations presented by \citet{2008MNRAS.386.1157C} through polynomials expressing a relationship between $A_{\text{FUV}}$ and the ratio of the total-infrared (TIR) and the $FUV$ luminosity. The advantage of using the $TIR$-to-$FUV$ ratio to determine the $FUV$ attenuation is the independence of the method on the relative dust-star geometry and the assumed extinction law \citep{1996A&A...306...61B,1999ApJ...521...64M,2000ApJ...533..236G,2000ApJ...528..799W}.
The total-infrared luminosity ($L_{\text{TIR}}$) on spatially resolved scales in M\,51 is computed from the MIPS\,24, PACS\,70 and 160\,$\mu$m maps (convolved to the PACS\,160\,$\mu$m resolution and rebinned to a grid with pixel size of 2$\arcsec$) and the calibrations presented by \citet{2013MNRAS.431.1956G}. The regression coefficients from \citet{2008MNRAS.386.1157C} are chosen as a function of the specific star formation rate (sSFR), as traced by different colors, to account for differences in the mean age of the stellar population, which will affect the calculation of $A_{\text{FUV}}$\footnote{The $A_{\text{FUV}}$ calibrations presented in \citet{2008MNRAS.386.1157C} were derived for a \citet{1955ApJ...121..161S} IMF, but have been shown to be appropriate for other initial mass functions (e.g. \citealt{2003PASP..115..763C}) only resulting in small variations of $A_{\text{FUV}}$ (0.1-0.2 dex) within the error bars.}. For M\,51, we constrain the age dependence based on $FUV-H$ colors (see Figure \ref{plot_Cortese}, left panel). The lowest $FUV-H$ colors correspond to high sSFRs, which are mainly found in the outer spiral arm regions located on the nearside of the interaction with NGC\,5195. Similarly, \citet{2010AJ....140..379K} found an enhancement in the number of stellar associations in the northern spiral arm closest to the companion, which suggest the triggering of star formation by the interaction.
The right panel of Figure \ref{plot_Cortese} shows the $A_{\text{FUV}}$ map derived following the procedures in \citet{2008MNRAS.386.1157C}. 
The $FUV$ attenuation clearly peaks in the center of M\,51, but also the spiral arm regions can be identified as regions of high extinction. The PACS\,70\,$\mu$m image contours correspond reasonably well with the observed features in the $FUV$ attenuation map. However, most of the extinction occurs in the inner regions of the spiral arms, which either suggests that most of the material is located on the inside of the spiral arm or, alternatively, that the star-forming regions are more obscured there. 

This $FUV$ attenuation map is used to constrain the geometrical distribution of dust by de-projection to a 3D geometry and assuming an exponential distribution in the $z$ direction. The exponential distribution of the dust has been proven adequate to describe the vertical distribution of dust in a galaxy's disk (e.g. \citealt{2013A&A...556A..54V,2014A&A...565A...4H}). 
The effect of a two-phase dust medium with clumpy dust features is studied and discussed in Section \ref{effect_clumps}, which shows that including the clumping of dust (in addition to the 2D asymmetric geometry derived from the $FUV$ attenuation map) causes an underestimation of the $FUV$ attenuation. We, therefore, argue that an additional component of quiescent compact dust clouds or clumps is not necessary to describe the dust component in the RT model of M\,51.

The resolution of the radiative transfer simulations is set by the size of individual grid cells, i.e. the physical area with constant dust and stellar mass fractions and properties. We, therefore, need to make sure to sufficiently sample the different regions within the galaxy to obtain reliable results from the RT calculations. We use a k-d tree grid structure, following the prescriptions from \citet{2014A&A...561A..77S}.
The $x \times y \times z$ dimensions of the grid are $20 \times 20 \times 5$ kpc$^{3}$ and the size of individual cells range from 40$^{3}$ to 312$^{3}$ pc$^{3}$. With an average optical depth $\tau_{\text{V}}$ $\sim$ 0.09 and maximum value $\tau_{\text{V}}$ $\sim$ 0.33 for the total number of grid cells (1,550,066), we are confident that the dust grid is well sampled and will produce reliable results simulating the interaction between stellar light and dust.

\begin{table}
\caption{Overview of the different stellar and dust components in the RT model of M\,51.}
\label{SKIRTcomp}
\centering
\begin{tabular}{l|l|c}
\hline  \hline 
Component & Parameter & Value \\
\hline \hline 
\multicolumn{3}{c}{Old stars} \\
\hline \hline 
Bulge\tablefootmark{a} & $n$ & 0.67 \\
	  & $R_{\text{e}}$ [pc] & 635.3 \\
	  & $q$ & 0.88 \\
	  & $L_{\text{V}}$ [$L_{\odot,\text{V}}$] & $3.2\times10^{9}$ \\
\hline 
Thick disk\tablefootmark{b} & 2D geometry & IRAC\,3.6\,$\mu$m\tablefootmark{c} \\
	          & $h_{\text{z}}$ [pc] & 450 \\
	         & $L_{\text{V}}$ [$L_{\odot,\text{V}}$] & $2.0\times10^{10}$ \\
\hline \hline 
\multicolumn{3}{c}{Young stars (non-ionizing)} \\ 
\hline \hline 
Thin disk\tablefootmark{d} & 2D geometry & \textit{GALEX} $FUV$\tablefootmark{e} \\
	          & $h_{\text{z}}$ [pc] & 100 \\
	          & SFR [M$_{\odot}$ yr$^{-1}$] & 3 \\
\hline \hline 
\multicolumn{3}{c}{Young stars (ionizing)} \\ 
\hline \hline 
Thin disk\tablefootmark{f} & 2D geometry & H$\alpha$ + 0.031 $\times$ MIPS 24\,$\mu$m\tablefootmark{g} \\
	          & $h_{\text{z}}$ [pc] & 100 \\
	         & $SFR$ [M$_{\odot}$ yr$^{-1}$] & 3 \\
	         & $M_{\text{d}}$ [$M_{\odot}$] & 4.5 $\times$ $10^6$ \\
 \hline \hline 
\multicolumn{3}{c}{Dust} \\ 
\hline \hline 
Thin disk\tablefootmark{h} & 2D geometry & $A_{\text{FUV}}$\tablefootmark{i} \\
	          & $h_{\text{z}}$ [pc] & 225 \\
	         & $M_{\text{d}}$ [$M_{\odot}$] & 7.3 $\times$ $10^7$ \\
\hline 
\end{tabular}
\tablefoot{
\tablefoottext{a}{The bulge component is modeled as a flattened S{\'e}rsic parametrized through the effective radius $R_{\text{e}}$, S{\'e}rsic index $n$, intrinsic flattening $q$ and $V$ band luminosity $L_{\text{V}}$.}\\
\tablefoottext{b}{The thick disk is modeled using observations to constrain the 2D geometry and assuming an exponential distribution with scale height $h_{\text{z}}$ in the vertical direction. The $V$ band luminosity, $L_{\text{V}}$, determines the normalization of the stellar emission spectrum.}\\
\tablefoottext{c}{Corrected for the bulge emission.}\\
\tablefoottext{d}{The thin disk of non-ionizing stars is modeled using observations to constrain the 2D geometry and assuming an exponential distribution with scale height $h_{\text{z}}$ in the vertical direction. The star formation rate, SFR, determines the normalization of the stellar emission spectrum.}\\
\tablefoottext{e}{Observed $FUV$ emission, not corrected for internal dust attenuation.}\\
\tablefoottext{f}{The thin disk of ionizing stars is modeled using observations to constrain the 2D geometry and assuming an exponential distribution with scale height $h_{\text{z}}$ in the vertical direction. The star formation rate, SFR, determines the normalization of the stellar+PDR emission spectrum. The dust mass, $M_{\text{d}}$, signifies the amount of grains present in the shells around young stellar clusters, associated with the PDR models.}\\
\tablefoottext{g}{The MIPS 24\,$\mu$m image is corrected for the emission of a diffuse dust component.}\\
\tablefoottext{h}{The thin dust disk is modeled using observations to constrain the 2D geometry and assuming an exponential distribution with scale height $h_{\text{z}}$ in the vertical direction. The dust mass, $M_{\text{d}}$, is used to normalize the amount of grains in the RT model.}\\
\tablefoottext{i}{Constructed based on the prescriptions from \citet{2008MNRAS.386.1157C}.}}
\end{table}

\subsubsection{SKIRT model spectra}
\label{Spectra.sec}

\textbf{Old stellar population} \\
We model the stellar emission spectrum of the old stars based on \citet{1998MNRAS.300..872M,2005MNRAS.362..799M} single stellar populations (SSP). The age and metallicity of the SSP are fixed to 10 Gyr old and $Z = 0.02$, consistent with the age and metallicity estimates in M\,51 reported by \citet{2004ApJ...615..228B,2010ApJS..190..233M} based on observational constraints and the fitting of stellar SEDs by \citet{2012ApJ...755..165M}.

\textbf{Young non-ionizing stars} \\
The stellar emission spectrum of $FUV$-emitting stars is described through Starburst 99 \citep{1999ApJS..123....3L} templates for an instantaneous starburst with solar metallicity ($Z=0.02$) and \citet{2002Sci...295...82K} IMF, normalized to a total mass of 10$^{6}$ M$_{\odot}$ and averaged over stellar ages between 10 and 100 Myr. 

\textbf{Young ionizing stars} \\
We describe the emission of ionizing stars based on the starburst templates from the library of pan-spectral SED models for young star clusters with ages $< 10$ Myr presented by \citet{2008ApJS..176..438G}. The SED templates of \citet{2008ApJS..176..438G} are based on a one-dimensional dynamical evolution model of H{\sc{ii}} regions around massive clusters of young stars combined with the stellar spectral synthesis code Starburst 99 \citep{1999ApJS..123....3L} and the nebular modeling code MAPPINGS III \citep{Groves2004}, assuming a \citet{2002Sci...295...82K} broken power-law IMF between 0.1 and 120 M$_{\odot}$.
The important parameters controlling the shape of the emission spectrum are the metallicity ($Z$) of the gas, mean cluster mass ($M_{\text{cl}}$), age and compactness ($C$) of the stellar clusters, the pressure of the surrounding ISM ($P_{0}$) and the cloud covering fraction ($f_{\text{PDR}}$). We choose a solar metallicity ($Z_{\odot}$) consistent with the metallicity of the old stellar component in the RT model of M\,51. The age parameter can be eliminated by averaging the spectra for all 21 cluster ages from 0.01 to 10 Myr. We assume fixed values for the mean cluster mass ($M_{\text{cl}}$ $\sim$ $10^{5}$ M$_{\odot}$) and ISM pressure ($P_{0}/k$ $\sim$ $10^{6}$ cm$^{-3}$ K). The latter approximations are based on the fact that more massive star clusters can be modeled as the superposition of several individual clusters and that the variation in ISM pressure mainly affects the nebular emission lines rather than the shape of the emission spectrum \citep{2008ApJS..176..438G}. We assume a cloud covering fraction $f_{\text{PDR}} = 0.2$, which is consistent with the assumptions of \citet{2010MNRAS.403...17J} applied for the modeling of star-forming regions in nearby galaxy samples and in line with observations in M\,51 and other nearby galaxies finding negligible fractions of completely obscured star-forming regions \citep{2005ApJ...633..871C,2007ApJ...668..182P}. Given our choices of the mean cluster mass $M_{\text{cl}}$ and ISM pressure $P_{0}/k$, we can choose values for the compactness parameter ranging from $\log C = 4.5$ to $6.5$. The compactness of the stellar clusters will mainly affect the temperature of dust grains with star clusters of lower densities having their peak at longer far-infrared wavelengths and being, thus, characterized by colder dust temperatures. We assume a moderate compactness factor of log $C$ = 5.5 (see also \citealt{2012MNRAS.427.2797D}), corresponding more or less to a warm dust component with dust temperature $T_{\text{d}} = 30 K$. 

\textbf{Dust component} \\
We assume an uniform composition of dust particles with a fixed grain size distribution throughout the galaxy, with the abundances of the grains (graphites, silicates and PAHs), extinction and emissivity of the dust mixture taken from the \citet{2007ApJ...657..810D} model derived for the dust in the Milky Way. 

\subsubsection{SKIRT model normalization}
\label{Normalization.sec}
Sections \ref{Geometry.sec} and \ref{Spectra.sec} have described our assumptions on the geometrical distribution of stars and dust, and the emission spectra for different stellar populations and dust properties, respectively. The normalization factors of stellar and dust components are the only free parameters in the model of M\,51, which are not fixed a priori but determined through minimization procedures.
The normalization of stellar components includes the scaling of the stellar emission spectra calculated for a single star to a certain bolometric or monochromatic luminosity representative for the stellar population. The luminosity of a single stellar photon will be determined as the total luminosity of the stellar component divided by the number of photon packages. The RT calculations presented in this paper are based on 5 $\times$ 10$^{6}$ photon packages. The normalization of the dust component sets the amount of dust particles in the RT calculations through the total dust mass.
Given that the density variations in the stellar and dust geometries are based on the relative intensity variations as perceived from observations, we rely on global photometry measurements to determine the normalization factors. 

Since the observed images used as input geometries for the RT calculations were cut at a certain signal-to-noise level, we might miss some of the extended stellar and dust emission observed in M\,51. Global photometry measurements for M\,51 published in the literature will, therefore, not be applicable for the RT model. We determine flux measurements adjusted to the coverage in the RT model of M\,51 by summing the emission within the areas of observed images that were retained as input geometries to the RT simulations.
The flux measurements derived in this manner are on average about $\sim$ 10-20$\%$ lower compared to published values in the literature that account for more extended emission features. 
The previously published photometry of M\,51 is likely to have been derived through different aperture photometry techniques from images to which different data reduction techniques and calibrations were applied. Given that our fluxes in all bands are determined in a homogeneous way, small differences between our photometry and the published photometry are expected. 

The uncertainties on the flux measurements include calibration uncertainties and stochastic uncertainties which reflect the uncertainties due to the method of flux extraction (see \citealt{2012A&A...543A.161C} for a more detailed description on error determination). The stochastic uncertainties include the instrumental uncertainties (related to the data coverage for each pixel), the background uncertainties and confusion noise. The background uncertainties are determined as the standard deviation of the mean values of several background regions (see Section \ref{process.sec}). The confusion errors only become significant in SPIRE wavebands, for which we rely on the confusion noise measurements from \citet{2010A&A...518L...5N}. The different error contributions are added in quadrature. The obtained fluxes are not corrected for the variation in beam size according to the shape of the spectrum. To allow a direct comparison between the observed band fluxes and the radiative transfer models, we convolve the output SED from the RT simulations with the appropriate response curves to obtain the model flux densities for a certain waveband.

The normalization constraints used for every RT component are briefly outlined. All model parameters are optimized in such a way that they reproduce the observed global flux densities within the error bars.
Table \ref{SKIRTcomp} provides an overview of the normalization factors ($V$ band luminosity $L_{V}$ for the old stellar populations in the bulge and thick disk, star formation rates $SFR$ for the (non-)ionizing young stars in the thin disk, and dust mass $M_{\text{d}}$ for the dust component in the thin disk) obtained from the model fitting described below.

\textbf{Old stellar population} \\
We constrain the luminosity of the old stars in the bulge and thick disk through the observed stellar SED. The global intensity of the old stellar component is scaled linearly until our models achieve the best fit with the stellar SED using a $\chi^2$ minimization procedure.
To constrain the observed stellar SED, we only account for the 2MASS $JHK$ and IRAC\,3.6\,$\mu$m wavebands since their emission is dominated by old stars and the effect of dust attenuation can safely be considered negligible. We scale the relative contributions to the global luminosity of the bulge and thick disk stellar components assuming a bulge-to-disk ratio of 0.16 (see fitting results from \citealt{2010AJ....139.2097P}). 

\textbf{Young stars} \\
The luminosity of the non-ionizing and ionizing stars is scaled to reproduce the observed global $FUV$ and MIPS\,24\,$\mu$m emission, respectively. We, hereby, postulate that the scaling of the input emission spectra of non-ionizing and ionizing stars correspond to the same star formation rate.

We obtain a SFR of $\sim$ 3 M$_{\odot}$ yr$^{-1}$ for young stars with ages younger than 100 Myr. The level of star formation obtained from the RT simulations corresponds well to the SFR estimates SFR(24\,$\mu$m) $\sim$ 3.4 M$_{\odot}$ yr$^{-1}$ and SFR(FUV) $\sim$ 4.3 M$_{\odot}$ yr$^{-1}$ derived by \citet{2005ApJ...633..871C} for a \citet{2001MNRAS.322..231K} IMF (consistent with a \citet{2002Sci...295...82K} IMF assumed in this work).

\textbf{Dust component} \\
The dust mass is scaled to reproduce the FUV attenuation. Since the $A_{\text{FUV}}$ map derived from the SSP fitting procedure varies throughout M\,51 and, thus, sets multiple constraints to the dust component in the radiative transfer model, we perform a least squares fitting procedure that minimizes the function:
\begin{equation}
f(\alpha) = \sum_{i=0,Npix} \left( \frac{A_{\text{FUV,obs}}[i] - A_{\text{FUV,RT}}[i]}{\sigma_{A_{\text{FUV,obs}}}[i]} \right)^{2}
\end{equation} 
which sums over all the pixels that resulted in reliable $A_{\text{FUV}}$ estimates. Each term in the sum is weighted by the uncertainty on the $FUV$ attenuation. 
The model $FUV$ attenuation is derived based on the output $FUV$ images for the RT simulations with and without dust: $A_{\text{FUV}}$ = -2.5 $\log$ $\left(I_{\text{total}}/I_{\text{transparent}} \right)$. To this, we add the $FUV$ attenuation derived for the young star clusters with ages $<$ 10 Myr based on the model properties for the emission spectra presented in \citet{2008ApJS..176..438G}. Given that the emission of photo-dissociation regions heated by young stars is modeled as a dusty foreground screen by \citet{2008ApJS..176..438G}, we convert the dust mass of young stellar clusters to a $FUV$ attenuation relying on the $V$ band attenuation-to-hydrogen column density ratio, $A_{\text{V}}$/$N_{\text{H}}$ = 5.34 $\times$ 10$^{-22}$ mag cm$^{2}$ H$^{-1}$, and the dust-to-hydrogen ratio, $\Sigma M_{\text{d}}/(N_{\text{H}} m_{\text{H}})$ =0.01, derived for the \citet{2007ApJ...657..810D} dust model. The dust attenuation curve applied by \citet{2008ApJS..176..438G} corresponds to the $R_{\text{V}}$ = $A_{\text{V}}$/$E_{\text{B-V}}$ = 3.1 curve presented by \citet{2005ApJ...619..340F} with $A_{\text{FUV}}$/$A_{\text{V}}$ $\sim$ 3, which is used to convert the obtained $A_{\text{V}}$ values into $FUV$ band attenuation.

The optimization of the normalization of young stellar emission and dust mass is performed simultaneously because the youngest stars are most obscured being still (partly) embedded in their birth clouds. The youngest stars are also hotter and emit the bulk of their energy at shorter wavelengths, which are more easily absorbed and scattered by dust grains. Based on the global $FUV$ and 24\,$\mu$m flux density of M\,51 and the $FUV$ band attenuation obtained from the procedure outlined in \citet{2008MNRAS.386.1157C} (see Section \ref{Dustcomp.sec}), we are able to simultaneously constrain the luminosity of young stars and the dust mass. A possible degeneracy between both parameters is non existent because an increase in dust mass will lower the $FUV$ emission and, therefore, require an additional scaling of the luminosity of young stars. The dust mass can, however, only be scaled to a maximum value that still reproduces the total infrared emission observed in M\,51. Within the uncertainties of the assumed dust properties, the resulting dust masses and stellar luminosities might differ depending on the variation in dust opacities and grain composition in the dust model. We assume an uniform dust population with grain composition and, extinction and emissivity properties for the dust mixture taken from the \citet{2007ApJ...657..810D} model for the dust in the Milky Way, which has been shown to reproduce the observed IR emission of several nearby galaxy samples (e.g. \citealt{2012ApJ...745...95D,2014A&A...565A.128C}).

Other than the dust mass used in the radiative transfer simulations, we have to account for the dust mass residing in the photo-dissociation regions surrounding H{\sc{ii}} regions. With the emission of individual stars clusters and their embedding dusty envelopes occurring on scales smaller than the resolution of our dust grid (i.e. 40 pc in the highest resolution dust grid cells), we have approximated their emission based on the library of SEDs presented by \citet{2008ApJS..176..438G} to model an ensemble of H{\sc{ii}} regions surrounded by photo-dissociation regions with a certain covering factor (here $f_{\text{PDR}}$ $\sim$ 0.2). The total dust mass associated with the young star clusters of ages $<$ 10 Myr, however, only accounts for 6 $\%$ of the total mass budget in M\,51. The two-phase dust component in the disk accounts for 7.3 $\times$ 10$^7$ M$_{\odot}$, while 0.4 $\times$ 10$^7$ M$_{\odot}$ is distributed in the outer shells of young star clusters. We find a total dust mass of 7.7 $\times$ $10^7$ M$_{\odot}$ which is about $\sim$ 35$\%$ lower compared to the dust mass $M_{\text{d}}$ $\sim$ 1.19 $\times$ 10$^8$ M$_{\odot}$ obtained from the dust SED modeling with the \citet{2007ApJ...657..810D} models reported by \citet{2012ApJ...755..165M}.  

The 35$\%$ difference in dust mass for M\,51 derived from the RT model (this paper) and a single modified-blackbody (MBB) fitting procedure \citep{2012ApJ...755..165M} can likely be attributed to the different model assumptions. The RT model, first of all, accounts for a wide range of dust temperatures along the line-of-sight (with single dust temperatures derived for every dust grain species within every 3D dust grid cell) as opposed to the single dust temperature approximation assumed for MBB fitting. The RT model is, furthermore, capable of self-consistently explaining the UV and mid-infrared emission from young stars, the dust obscuration and emission, which provides strong constraints on the sources of dust heating (see Section \ref{Analysis.sec}) and relative dust-star geometry (see Section \ref{effect_scaleheight}).  
Based on the dust mass derivation from the RT model ($M_{\text{d}}$ = 7.7 $\times$ 10$^7$ M$_{\odot}$) and the gas mass reported in \citet{2012ApJ...755..165M}, we obtain a gas-to-dust mass ratio of 142, which is consistent with the gas-to-dust mass fractions obtained for our Milky Way (160, \citealt{2004ApJS..152..211Z}) and other nearby late-type spiral galaxies from the \textit{Herschel} Reference Survey \citep{2012A&A...540A..52C}. 

\section{RT model validation}
\label{Fitting.sec}

\subsection{Global galaxy scale}

\begin{figure*}
\centering
\includegraphics[width=18.5cm]{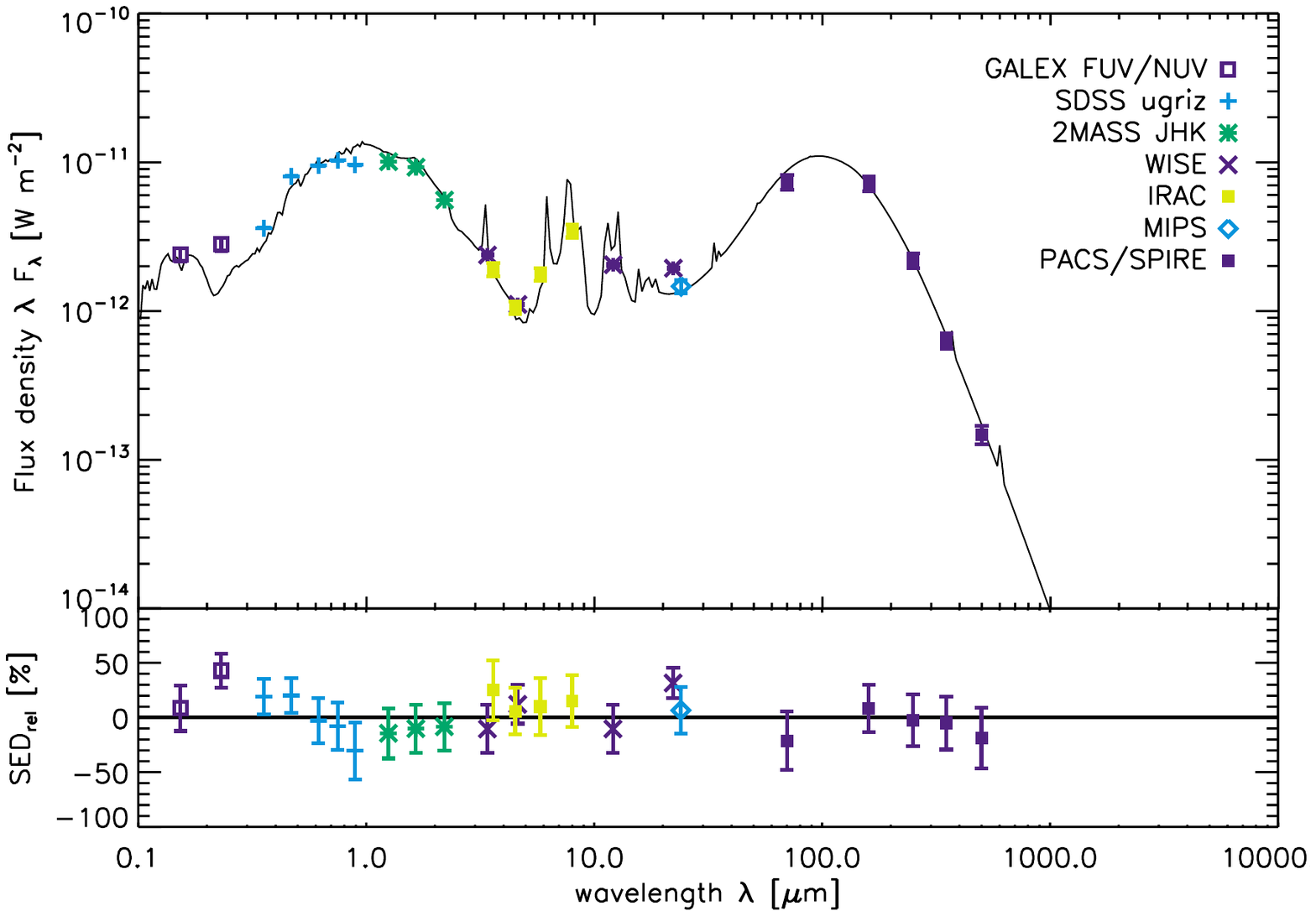}   
 \caption{Top: the multi-wavelength spectral energy distribution of M\,51 where the black solid line represents the emission derived from the radiative transfer model for M\,51 (including dust, old and young stars) and the legend indicates the origin of the different data measurements. Bottom: the residual SED obtained as the relative difference between observations and model flux densities or SED$_{\text{rel}}$[$\%$] = ($F_{\text{obs}}$-$F_{\text{model}}$)/$F_{\text{obs}}$.}
              \label{plot_step3_SED}
\end{figure*}

Figure \ref{plot_step3_SED} shows the multi-wavelength SED in our radiative transfer model (black solid line) overlaid with the observed multi-waveband constraints, derived through summing the emission within the high signal-to-noise regions in M\,51 (see Section \ref{Normalization.sec}). Figure \ref{plot_step3_SED} indicates that the RT model generally reproduces the observed far-infrared/submillimeter emission in M\,51 within the error bars. The amount of absorbed stellar energy (i.e. 10$\%$ of the total stellar bolometric luminosity) is, thus, shown to be in perfect balance with the thermal energy re-emitted by dust grains. The dust energy balance in M\,51 is, hereby, in contrast with conclusions of many similar studies on edge-on spiral galaxies \citep{2000A&A...362..138P,2001A&A...372..775M,2004A&A...425..109A,2005A&A...437..447D,2010A&A...518L..39B,2011A&A...527A.109P,2012MNRAS.419..895D,2012MNRAS.427.2797D,2012A&A...541L...5H} reporting the underestimation of the observed FIR/submm emission by a factor of 3-4 in RT models, which were constrained based on optical observations of the dust attenuation.
Different scenarios were invoked to reconcile the model calculations with the observations, among which higher FIR/submm dust emissivities \citep{2004A&A...425..109A,2005A&A...437..447D,2011ApJ...741....6M} or the distribution of a major dust fraction in geometries (second thin inner dust disk, two-phase clumpy dust) that have a negligible effect on the dust attenuation \citep{2000A&A...362..138P,2001A&A...372..775M,2008A&A...490..461B,2011A&A...527A.109P,2011ApJ...741....6M} are the most favorable explanations. 

We believe that the reason for the dust energy balance in M\,51 compared to the inconsistency in the dust energy budget for edge-on galaxies can be looked for in the different aspects of our modeling approach. The main difference in construction of our RT model depends on the viewing angle of M\,51, which allows to discern the position of star-forming regions and asymmetries in stellar and dust geometrical distributions. The adequate description of these sites of young star clusters and the observed asymmetric structures in the RT model allows to make a careful prediction of the radiative heating of dust by young and old stellar populations, resulting in a more reliable estimate of the thermal dust emission at IR/submm wavelengths. Another difference lies in the dust geometry and the constraints used to scale the total dust mass in the RT model. Rather than constraining the dust attenuation from optical observations, we have calculated the $FUV$ attenuation based on the \citet{2008MNRAS.386.1157C} recipes which simultaneously rely on $FUV$ and infrared observations to derive $A_{\text{FUV}}$ estimates. The compatibility between the absorbed energy of stars and the re-emitted dust emission in M\,51, diverging from what is observed for edge-on galaxies, suggests that the origin of the dust energy balance problem in edge-ons results from the incapability of constraining the location and emission of star-forming regions and determining the amount of dust attenuation solely from UV/optical observations (see also \citealt{2013ApJ...771...62K}). The resolution of optical data in the determination of dust attenuation has also been shown to be of great importance (e.g. \citealt{2013ApJ...778L..41L}) due to the small spatial scales (few parsecs) at which most of the attenuation processes take place. 

To better highlight the differences between the observed and model emission in several wavebands, we show the residual flux densities, i.e. relative difference between observations and model flux densities or SED$_{\text{rel}}$[$\%$] = ($F_{\text{obs}}$-$F_{\text{model}}$)/$F_{\text{obs}}$, in the bottom panel of Figure \ref{plot_step3_SED}. The uncertainties on the residuals include the uncertainties on the observed flux densities as well as the Monte Carlo noise from the radiative transfer calculations. The modeled SKIRT fluxes -distributed in a wavelength grid with 500 steps ranging from 0.05 to 1000 $\mu$m- are convolved with the response curves of the respective bands to obtain the model flux. 

\begin{figure}
\centering
\includegraphics[width=9.0cm]{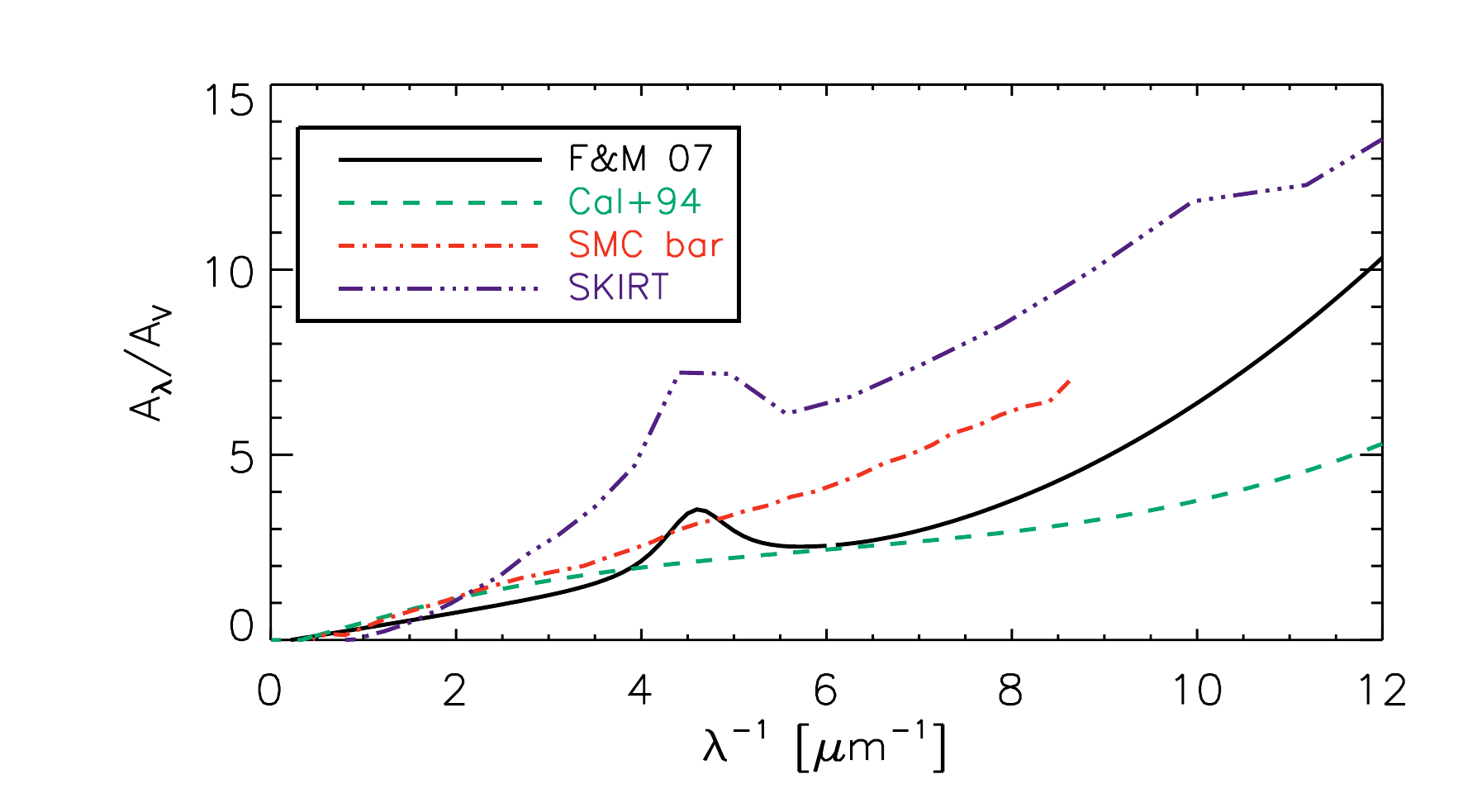}   
\caption{The dust attenuation curve as derived from our radiative transfer calculations (triple dotted-dashed purple line). The dust extinction curves derived for the Milky Way with $R_{\text{V}}$ = 3.0 \citep{2007ApJ...663..320F} and the Small Magellanic Clouds \citep{2003ApJ...594..279G} and the \citet{1994ApJ...429..582C} attenuation law for starburst galaxies are added for comparison as black solid, red dotted-dashed and green dashed lines, respectively.}
              \label{plot_SKIRTattcurve}
\end{figure}

In most wavebands, the observed and modeled fluxes agree within the error bars. One clear exception is the observed emission in the $NUV$ band which is significantly lower in the RT model compared to the observations. We argue that the discrepancy can be attributed, at least in part, to the dust bump at 2175$\AA$ characterizing the dust attenuation curve of our RT model for M\,51. The output dust attenuation curve from the RT model (see Figure \ref{plot_SKIRTattcurve}) shows a similar dust absorption feature as observed in the Galactic dust attenuation curve with selective-to-total extinction $R_{\text{V}}$ = 3.0 \citep{2007ApJ...663..320F}. 
Based on the analysis of \citet{2005ApJ...633..871C}, the dust attenuation in M\,51 seems characterized by a weak 2175$\AA$ absorption feature, as compared to the clear 2175$\AA$ dust bump in the Milky Way (MW) extinction curve, and to better resemble the Small Magellanic Clouds (SMC) extinction curve \citep{2003ApJ...594..279G}. With the 2175$\AA$ dust feature being absent in the lower metallicity environments of the SMC and LMC \citep{2003ApJ...594..279G}, the metallicity of galaxies has been argued to play a role in the strength of the bump. With the metallicity of M\,51 being solar to even super-solar locally, we would expect a prominent 2175$\AA$ absorption feature if the metallicity plays a role in shaping the dust composition and, thus, the shape of the attenuation curve. Other than metallicity, the absorption bump at 2175$\AA$ appears to be sensitive to the UV radiation field (e.g. \citealt{2011MNRAS.417.1760W}) which can alter the dust grain size distribution through destruction/coagulation processes and/or change the ionization state of dust grains \citep{1988ApJ...332..328B,1996A&A...315L.305C}.  Also radiative transfer effects coupled with age-selective attenuation have been shown to affect the bump strength \citep{1998ApJ...509..103S,2000ApJ...542..710G}. The strength of the attenuation bump, furthermore, depends on the carriers of the extinction feature \citep{1997ApJ...487..625G,2000ApJ...528..799W} which could be related to dust grains composed of aromatic carbon \citep{1989IAUS..135..313D} such as graphite grains or polycyclic aromatic hydrocarbons (PAHs).
We, therefore, argue that the dust population in M\,51 might differ from the standard \citet{2007ApJ...657..810D} dust composition for the Milky Way. In Section \ref{effect_clumps}, we also show that the strength of the 2175$\AA$ absorption feature might be influenced by the clumping fraction of dust.

Other than the dust 2175$\AA$ absorption feature, the $UV$ slope in the output attenuation curve from the RT model appears significantly steeper compared to the extinction curves derived for the MW \citep{2007ApJ...663..320F}, SMC \citep{2003ApJ...594..279G} and a sample of starburst galaxies \citep{1994ApJ...429..582C}. We do need to caution that the uncertainty on the $V$ band attenuation of the RT model is large due to the relatively fewer $V$ band photons that are scattered or absorbed compared to the shorter wavelength UV photons (e.g. see the grainy appearance of the inter-arm regions in the $\tau_{\text{V}}$ map in Figure \ref{plot_step2_tau} due to Poisson noise). \textbf{The output attenuation curve from SKIRT, furthermore, depends on the intrinsic extinction curve for the dust grains (which is similar to the MW extinction curve), but also reflects any 3D geometrical and/or scattering effects.} If the curve is representative for the attenuation law of the RT model of M\,51, the strong $UV$ slope might be telling us that the total-to-selective extinction ratio, $R_{\text{V}}$, is smaller than the average $R_{\text{V}}$ = 3.1 observed in the Milky Way \citep{1989ApJ...345..245C}. The curve with $R_{\text{V}}$ = 3.1 for the Milky Way is, however, an extinction law which does not include the effects of scattering and geometry. The attenuation curve from SKIRT does account for these effects, which could explain part of the difference. Scattering and the relative dust-star geometry have indeed been shown to play an important role in characterizing the attenuation law in face-on galaxies \citep{2001MNRAS.326..733B}. \textbf{A further investigation of the SKIRT attenuation curve with different dust models and relative dust-star geometries is beyond the scope of the present study.}

\subsection{Spatially resolved galaxy scales}
\label{Spat.sec}
While the normalization of the SKIRT RT stellar and dust components (see Section \ref{Normalization.sec}) is optimized for global galaxy measurements (except for the $FUV$ attenuation), deviations from the observed multi-waveband images of M\,51 might occur in our radiative transfer model on spatially resolved scales. Even though the observed 2D structures in M\,51 are used to constrain the dust and stellar geometries in the model, we might have disregarded spatial variations in the age of stellar populations and/or deviations from a uniform dust composition throughout the galaxy. We need to investigate whether local dissimilarities occur between the radiative transfer model and observations and, more importantly, understand the shortcomings of our model that result in its inability to reproduce the observed characteristics. Therefore, we have convolved all observed and model images to the working resolution of $\sim$ 12.1$\arcsec$ (or 493 pc) in the PACS\,\,160\,$\mu$m waveband. 

\subsubsection{Stellar and dust emission}
\begin{figure*}
\centering
\includegraphics[width=18.5cm]{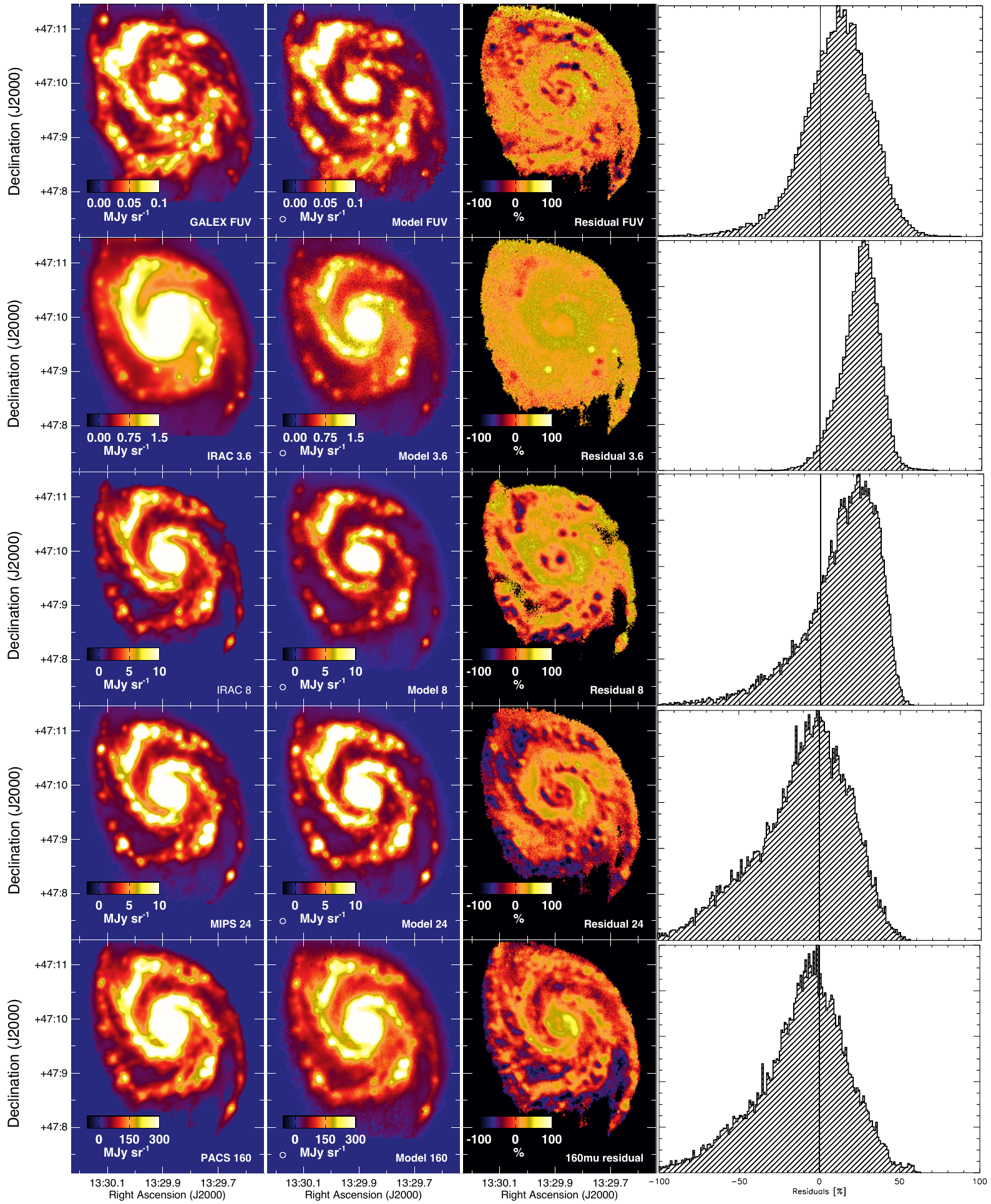}   
 \caption{The observed (left column), modeled (2nd column) and residual (3rd column) images of M\,51 for the \textit{GALEX} $FUV$ (top row), IRAC\,3.6\,$\mu$m (2nd row), IRAC\,8\,$\mu$m (3rd row), MIPS\,24\,$\mu$m (4th row) and PACS\,160\,$\mu$m (5th row) wavebands. The observed images have been convolved to the resolution of the PACS\,160\,$\mu$m filter, i.e. the resolution of all input parameters maps in the radiative transfer code. The size of the PACS beam ($\sim$ 12 arcsec) is indicated in the bottom left corner of the model images. The last column shows the histogram of residuals (normalized to 1 at the peak).}
             \label{plot_model_combined}
\end{figure*}

Figure \ref{plot_model_combined} compares a handful of observed images of M\,51 (1st column) with the model results from the RT calculations (2nd column). The third column shows the relative residuals, i.e. ($I_{\text{obs}}$ - $I_{\text{model}}$)/$I_{\text{obs}}$, with the histogram of residuals (normalized to 1 at the peak) shown in the last column. From top to bottom, we compare the images observed in \textit{GALEX} $FUV$, IRAC\,3.6\,$\mu$m, IRAC\,8\,$\mu$m, MIPS\,24\,$\mu$m and PACS\,160\,$\mu$m wavebands. 

The $FUV$ emission in the model matches reasonably well to the observed $FUV$ emission in M\,51 with the exception of some regions in the spiral arms, where the model overestimates the observed $FUV$ emission up to 60$\%$, and the underestimation of the FUV emission in the inter-arm regions.
These small variations between the model and observations are most likely a reflection of the stellar clusters of different ages that populate the regions of M\,51 throughout the centre and spiral arms. While the radiative transfer model assumes an averaging of ages for stars younger than 100 Myr throughout the galaxy plane, the stellar clusters of specific ages are more likely clumped together in different regions and not homogeneously spread throughout the galaxy (see for instance \citealt{2005ApJ...619L..71B,2010AJ....140..379K}). Darker regions in the residual map indicate areas where the emission of young ionizing stars ($<$ 10 Myr) is overestimated and might, thus, merely consistent of non-ionizing stars. Alternatively, the escape fraction of hard UV photons from their birth clouds might be overestimates in those regions. The bright regions in the residual $FUV$ map indicate that the RT model lacks a more diffuse emission component in the inter-arm regions, which could hint at the presence of a thick disk component with dust that scatters $UV$ photons into the line of sight. Alternatively, we might underestimate the escape fraction of young stellar photons which could produce a diffuse emission component.

The good correspondence of the modeled 3.6\,$\mu$m map with the observed IRAC\,3.6\,$\mu$m image (i.e. model and observations agree within 50$\%$) suggests that the old stellar component in the RT model is representative for the population of old stars in M\,51.
Small deviations can be perceived in the inter-arm regions of M\,51, where the model underestimates the diffuse stellar emission. 
The 3.6\,$\mu$m emission seems, furthermore, overestimated in some regions of the spiral arms, which suggests that the contribution of young stars to the near-infrared wavebands is somewhat too high in the model.

The IRAC\,8\,$\mu$m emission is overestimated in the spiral arms by at most 50$\%$, whereas the model also under-predicts the PAH emission observed in some inter-arm regions of M\,51. The absence of bright 8\,$\mu$m emission in the spiral arms might, partly, be linked to the underestimation of the 24\,$\mu$m emission and, thus, star formation activity (see later). Alternatively, the improper characterization of the PAH abundance in the RT model (assuming a constant PAH fraction throughout the galaxy), which does not account for the destruction of PAHs exposed to hard radiation fields (e.g. \citealt{2004A&A...428..409B,2004ApJS..154..253H,2005ApJ...633..871C,2007ApJ...663..866D,2007ApJ...668..182P,2008MNRAS.389..629B,2008ApJ...682..336G}), might explain the differences between model and observations. 

We observe a similar pattern in the 24\,$\mu$m emission of M\,51, with the model overestimating the emission up to 50$\%$ in the spiral arm structures, i.e. the sites of young stellar activity, and underestimating the diffuse 24\,$\mu$m in the inter-arm regions and disk of M\,51.
We argue that the former discrepancy could again originate from the uniform age range assumed for young stars throughout M\,51 (averaging all ages younger than 100 Myr), while radial differences have been shown to occur with a decrease in age of the young stellar populations towards larger radial distances from M\,51's centre \citep{2005ApJ...633..871C}. Alternatively, the deviation from an uniform dust grain composition in some parts of the galaxy could contribute to the spatial inaccuracies in our modeled 24\,$\mu$m image. Since the abundance of very small grains (VSGs) has been shown to trace the star formation activity (e.g. \citealt{2005ApJ...633..871C,2008AJ....136..919B,2009AJ....138..196P}), it would not be surprising if the abundances of very small grains would be higher in localized star-forming regions due to the shattering of larger grains. The increase of the PAH\,8\,$\mu$m-to-MIPS\,24\,$\mu$m ratio with galactocentric radius \citep{2013MNRAS.433.1837V}, indeed, suggests that the excitation and/or abundance of PAHs and very small grains is not uniform throughout the galaxy's disk.

At longer infrared wavebands, we observe the opposite trend with the model underestimating the emission from the spiral arms and overestimating the diffuse dust emission in the disk of M\,51 up to 50$\%$.

In summary, we find local variations in the RT model emission deviating from the observed stellar and dust emission in M\,51. The local deviations suggest that the model assumptions of an uniform dust composition and the averaging of stellar ages throughout the galaxy are only adequate to first order. The characterization of the dust grain size distribution, abundances and grain emissivities in combination with the stellar age variation requires several additional datasets that are sensitive to the emission and absorption features of individual grain populations (e.g. \textit{Spitzer} IRS, \textit{SWIFT}, ...) and is, therefore, beyond the scope of this current work. The underestimation of the diffuse emission in M\,51, furthermore, suggests that we might need to consider the presence of an additional dust component distributed in a disk with larger scale height. 

\subsubsection{Dust attenuation}
\begin{figure*}
\centering
\includegraphics[width=18.5cm]{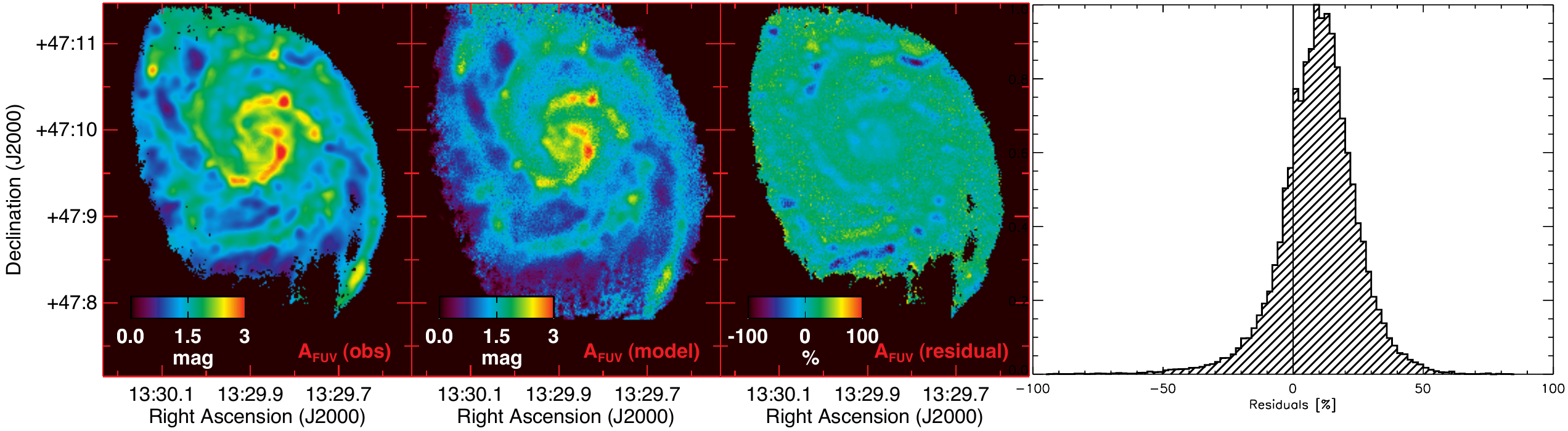}   
 \caption{From left to right: the observed, modeled and residual image of the $FUV$ dust attenuation, $A_{FUV}$. The last panel shows the histogram of residuals (normalized to 1 at the peak).}
             \label{plot_step2_AFUV}
\end{figure*}

Figure \ref{plot_step2_AFUV} presents the observed, model and residual images of the $FUV$ dust attenuation in M\,51 (see first three panels) with the histogram of residuals (normalized to 1 at the peak) shown in the last column.
The observed $FUV$ attenuation map derived from the procedures outlined by \citet{2008MNRAS.386.1157C} is well reproduced by our model, which suggests that the relative orientation of stars and dust in our model and, thus, the scale heights of dust and stars are modeled in a realistic way. 
Small deviations occur in some localized sources in the spiral arms which suggests that our model overestimates the attenuation in parts of the galaxy, mainly located in the outer regions of the spiral arms. With those outer spiral arms being mostly affected by the interaction with NGC\,5195,  the material in the outer arm regions might have been re-arranged due to the intervention of NGC\,5195, making $UV$ photons capable of escaping more easily from their birth clouds.
A metallicity gradient, which has not been accounted for in the RT model assuming a constant metallicity, might also cause the $FUV$ attenuation to be lower at longer galacto-centric distances.
The variations between model and observations, however, generally remain within the uncertainties of the $FUV$ attenuation.  

\subsubsection{Optical depth}
\begin{figure}
\centering
\includegraphics[width=8.5cm]{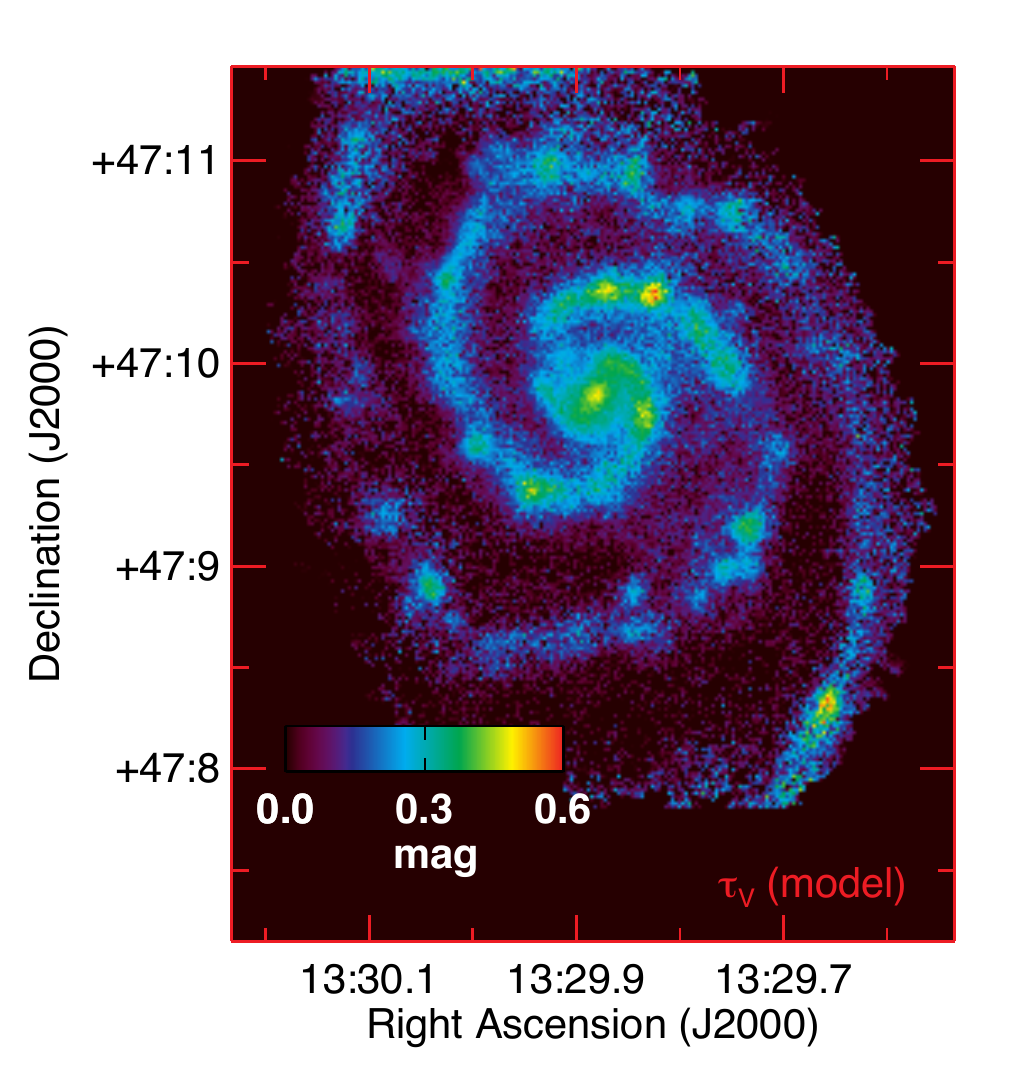}   
 \caption{The optical depth in the $V$ band, $\tau_{\text{V}}$, derived from the RT calculations for M\,51. The $\tau_{\text{V}}$ map has a grainy structure in the inter-arm regions due Poisson noise, driven by the relatively few number of $V$ band photons that are absorbed and/or scattered in the inter-arm regions.}
             \label{plot_step2_tau}
\end{figure}

To analyze the transparency of the stellar disk at the inclination of M\,51, we compare the RT model with and without dust by constructing a map of the optical depth, $\tau_{\text{V}}$ = $\exp$(-$I_{\text{V,total}}/I_{\text{V,transparent}}$), from the model (see Figure \ref{plot_step2_tau}), with $I_{\text{V,total}}$ being the total $V$ band intensity and $I_{\text{V, transparent}}$ the intrinsic $V$ band intensity in the absence of any dust (i.e. a transparent medium). For values $\tau_{\text{V}}$ $<$ 1, the galaxy is considered nearly transparent, implying that most of the stellar radiation can be observed directly without the obscuring effect of dust. Higher values of $\tau_{\text{V}}$ would imply that a significant portion of the stellar $V$ band light is absorbed and reprocessed by dust grains. Based on radiative transfer simulations of edge-on spiral galaxies, the face-on optical depth in most optical wavebands is generally concluded to be lower than one, making the disk nearly transparent in the optical wavelength regime \citep{1997A&A...325..135X,1998A&A...331..894X,1999A&A...344..868X,2004A&A...425..109A,2007A&A...471..765B,2011A&A...527A.109P,2013A&A...550A..74D,2014MNRAS.441..869D}. Some exceptions were found for the secondary dust disk used to explain the missing FIR emission in RT models of edge-on galaxies \citep{2000A&A...362..138P,2001A&A...372..775M,2011A&A...527A.109P,2011ApJ...741....6M}. The model fitting procedures of edge-on galaxies have, however, been shown to be influenced by model degeneracies between dust scale length and face-on optical depth \citep{2014MNRAS.441..869D}, which could hamper the derivation of a reliable $\tau_{\text{V}}$ estimate. 

The RT model for M\,51 shows a variation in optical depth from $\tau_{\text{V}}$ $<$ 0.1 in the inter-arm regions, $\tau_{\text{V}}$ $\sim$ 0.3 on average in the spiral arms with peaks up to $\tau_{\text{V}}$ $\sim$ 0.6 in localized star-forming regions in the spiral arms and the center of M\,51. The results for M\,51 support the earlier analyses of edge-on galaxies indicating that the optical stellar light of galaxies is hardly influenced by extinction when viewed face-on. The relatively small effects of dust in optical wavebands is re-assuring for the determination of intrinsic galaxy properties such as stellar mass, star formation history, star formation rate and global galaxy population diagnostics such as initial mass function, luminosity functions and colour-magnitude diagrams (e.g. \citealt{2007MNRAS.379.1022D,2009ApJ...691..394M,2010MNRAS.403.1894D}). We should, however, extend the same analysis to larger samples of galaxies with wide spread in star formation rate to confirm that also the optical disks of more actively star forming disks hardly experience effects of dust extinction. We should, furthermore, caution that the $\tau_{\text{V}}$ values are averaged on scales of 12.1$\arcsec$ (or 493 pc at the distance of M\,51), which does not exclude the possibility that higher optical depths occur on scales smaller than 500 pc (i.e. the size of typical giant molecular clouds). Averaged over larger areas, the peaks of high optical depth will be smoothed out (e.g. \citealt{2013ApJ...778L..41L}).

\subsubsection{Infrared colors}
\begin{figure*}
\centering
\includegraphics[width=18.5cm]{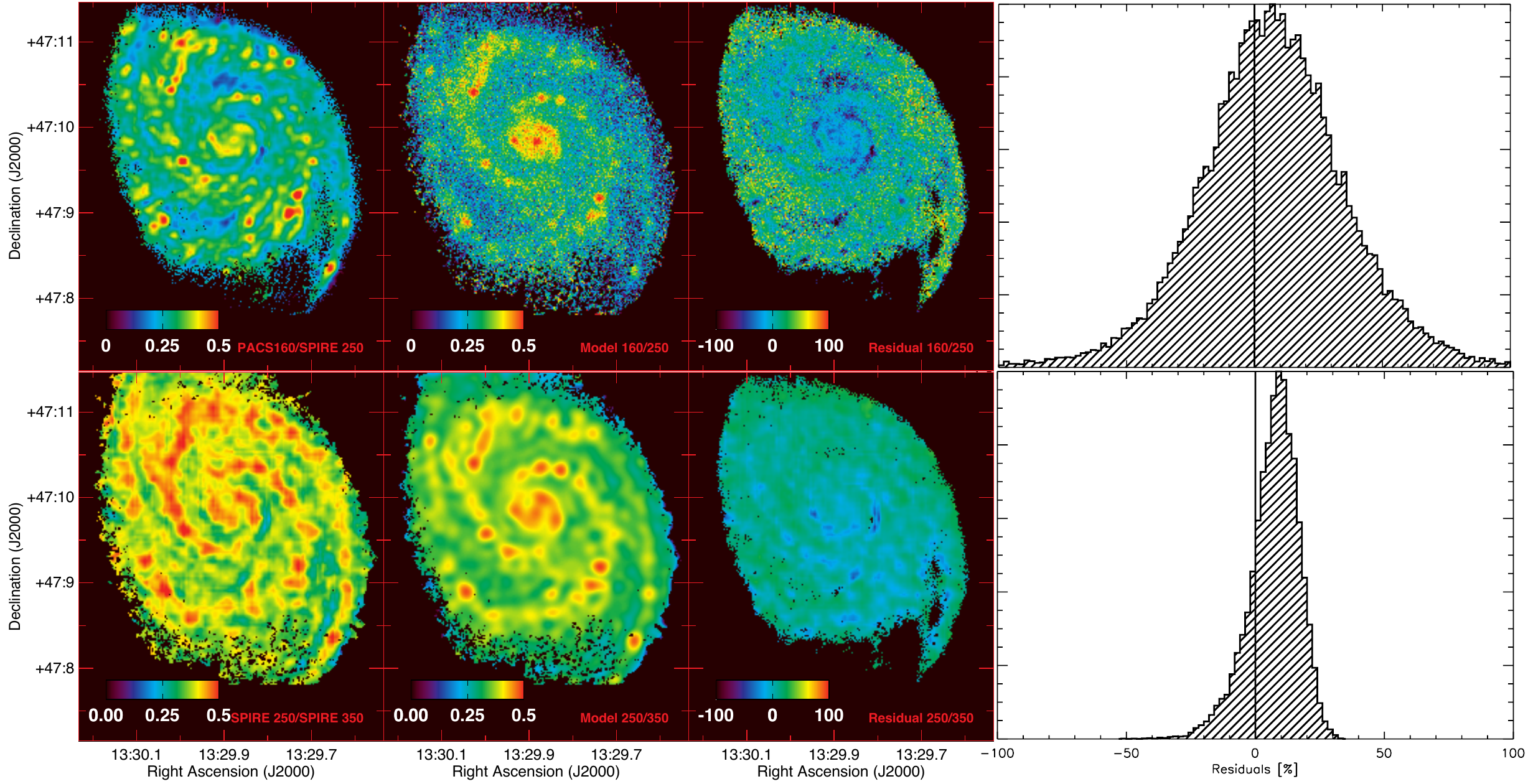}   
\caption{The surface brightness intensity ratio maps PACS\,160\,$\mu$m-to-SPIRE\,250\,$\mu$m (top) and SPIRE\,250\,$\mu$m-to-SPIRE\,350\,$\mu$m (bottom) as derived from observations (left panel) and radiative transfer calculations (second panel). The panels in the third and last row show the residual images and the histogram of residuals (normalized to 1 at the peak), respectively.}
             \label{plot_colour}
\end{figure*}

We verify whether our RT model is sensitive to the dust temperature variation on spatially resolved scales and, thus, capable to reproduce the observed infrared colors. We compare the flux density ratios, PACS\,160\,$\mu$m-to-SPIRE\,250\,$\mu$m and SPIRE\,250\,$\mu$m-to-SPIRE\,350\,$\mu$m, as derived from observations to the output from radiative transfer calculations (see Figure \ref{plot_colour}). If we assume that a single modified black-body provides an adequate description of the dust SED at wavelengths long wards of about 100\,$\mu$m (e.g. \citealt{2012ApJ...756...40S,2012MNRAS.425..763G,2013MNRAS.428.1880A,2013A&A...557A..95R,2014MNRAS.440..942C}), i.e.
\begin{equation}
F_{\nu}~=~\frac{M_{d}}{D^{2}} \kappa_{\nu_{0}} \left( \frac{\nu}{\nu_{0}} \right)^{\beta} B_{\nu}(T),
\end{equation}
then the ratio of the flux densities at different infrared wavebands will be independent of the dust absorption coefficient, $\kappa_{\nu_{0}}$, defined at a reference wavelength $\nu_{0}$.
Indeed, the flux ratios or infrared colors, i.e.
\begin{equation}
\frac{F_{\nu}(\nu_{1})}{F_{\nu}(\nu_{2})} = \left( \frac{\nu_{1}}{\nu_{2}} \right)^{\beta+3} \left( \frac{e^{h \nu_{2}/kT}-1}{e^{h \nu_{1}/kT}-1} \right),
\end{equation}
will only depend on the dust emissivity index, $\beta$, and the dust temperature, $T_{\text{d}}$.
Figure \ref{plot_colour} shows the observed (1st column), modeled (2nd column) and residual (3rd column) flux density ratios of PACS\,160\,$\mu$m-vs-SPIRE\,250\,$\mu$m (top) and SPIRE\,250\,$\mu$m-vs-SPIRE\,350\,$\mu$m (bottom). The last column shows the histogram of residuals, normalized to 1 at the peak. The spatial resemblance of peaks and hollows between the observed and modeled surface brightness intensity ratios makes us confident that the RT model is capable of giving an unbiased view on the importance and variance in dust heating sources across the galaxy. The PACS\,160\,$\mu$m-vs-SPIRE\,250\,$\mu$m color shows good global agreement with some local deviations from the observed ratios. The PACS\,160\,$\mu$m-vs-SPIRE\,250\,$\mu$m colors are too warm in some localized regions in the inter-arms of M\,51, which could indicate that the escape fraction of young stars from star-forming regions in the spiral arms is too high and capable of heating the diffuse dust in the inter-arm regions. The agreement between model and observations is better for the SPIRE\,250\,$\mu$m-to-SPIRE\,350\,$\mu$m color with a relatively small scatter in the distribution of residuals around the mean value, which suggests that the heating of the colder dust component in M\,51 is accounted for in a most reliable way by the RT model. 

\section{Dust heating in M\,51}
\label{Analysis.sec}

\subsection{Wavelength-dependent dust heating}
\begin{figure}
\centering
\includegraphics[width=8.5cm]{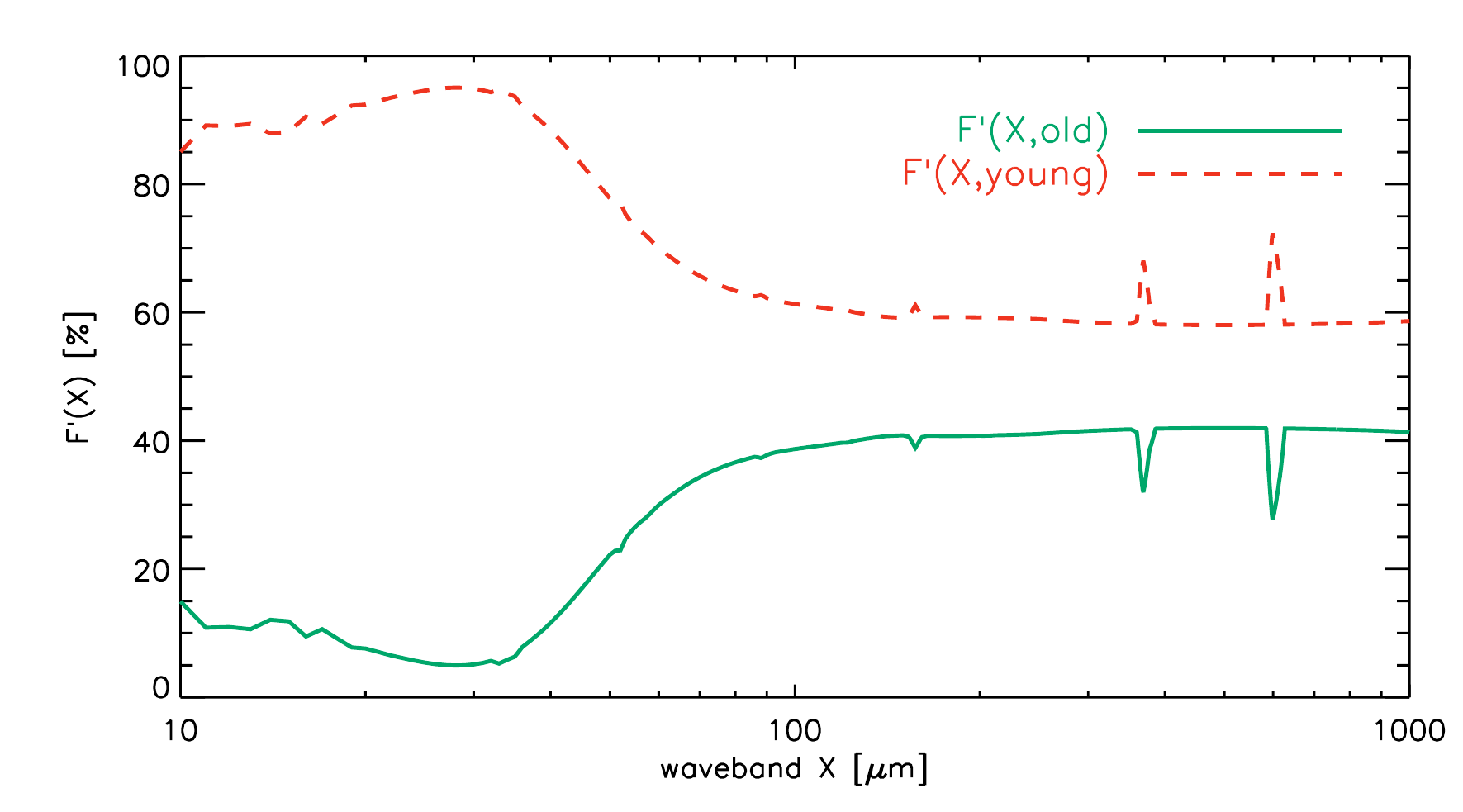}   
 \caption{The contribution from old (green solid line) and young stars (red dashed line) to the radiative heating of dust across the multi-wavelength infrared spectrum, as determined from the radiative transfer model of M\,51.}
             \label{plot_wave_dustheating}
\end{figure}

The 3D high-resolution radiative transfer model of M\,51 allows a self-consistent study of the dust heating mechanisms on spatial scales of 500 pc.
By tracing the individual photons of the different stellar populations in our radiative transfer calculations, we can investigate how much of the dust emitting at certain wavelengths is heated by the young and old stellar population. In the case of M\,51, we ignore possible AGN activity as a significant source of mid-infrared emission and, thus, dust heating based on the results from \citet{2012ApJ...755..165M}. We make a distinction here between the old stellar population and the radiation from young stars with ages $\lesssim$ 100 Myr. A clear distinction between the two sources of dust heating is hampered by the non-linearity of the relation between the shape of the absorbed emission spectrum and the re-emitted emission at specific wavelengths. The sum of the energy absorbed from young and old stars is equal to the total absorbed energy (i.e., $L_{\text{abs,total}}$ = $L_{\text{abs,old}}$ + $L_{\text{abs,young}}$) which can be translated into the total energy re-emitted at infrared and (sub-)millimeter wavebands (i.e., $L_{\text{TIR,total}}$ = $L_{\text{TIR,old}}$ + $L_{\text{TIR,young}}$). The absorbed energy of young and old stars, separately, can, however, not simply be translated in the dust emission at specific monochromatic wavebands. More specifically, the dust emission in a monochromatic waveband depends on the global embedding radiation field and, therefore, can not be separated into a young and old emission spectrum, i.e.: $F_{\text{X,total}} \neq F_{\text{X,old}} + F_{\text{X,young}}$, for a monochromatic waveband X. Since the non-linearity in the dependence of dust heating on the radiation field can not be eliminated from the RT calculations, the heating due to old and young stars cannot strictly be separated. To account for this non-linear behavior, we approximate the heating by old and young stars through
\begin{equation}
\label{eq1}
F'_{\text{X,old}} = \frac{0.5 F_{\text{X,old}} + 0.5 (F_{\text{X,total}} - F_{\text{X,young}})}{F_{\text{X,total}}}
\end{equation}
and,
\begin{equation}
\label{eq2}
F'_{\text{X,young}} = \frac{0.5 F_{\text{X,young}} + 0.5 (F_{\text{X,total}} - F_{\text{X,old}})}{F_{\text{X,total}}},
\end{equation}
respectively, for which $F'_{\text{X,total}}$ $=$ $F'_{\text{X,old}}$ $+$ $F'_{\text{X,young}}$ does hold. The flux densities $F_{\text{X,total}}$, $F_{\text{X,old}}$ and $F_{\text{X,young}}$ are derived from repeated radiative transfer simulations with a stellar population consisting of old+young, old and young stars, respectively. 
Figure \ref{plot_wave_dustheating} shows the relative fraction of energy from old and young stars contributing to the heating of dust emitting at infrared wavelengths from 10 to 1000\,$\mu$m. Table \ref{table_dustheating} indicates the fraction (in $\%$) of dust heating by old stars (middle column) and young stars (right column) at several wavebands. Based on the radiative transfer calculations, we can conclude that the dust heating by the young stars in M\,51 clearly dominates in mid-infrared wavebands and gradually decreases towards longer wavelengths. The evolved stellar population has a non-negligible contribution to the dust heating in far-infrared and submillimeter wavebands, which is consistent with previous multi-band studies of the dust heating instigators (e.g. \citealt{2008A&A...490..461B,2010A&A...518L..65B,2011AJ....142..111B,2011A&A...527A.109P,2012MNRAS.419.1833B,2012A&A...543A..74X,Bendo}). 
Nonetheless, the stellar energy provided by young stars remains the dominant heating source also at longer wavelengths. The maximum contribution of $\sim$ 40$\%$ from the old stellar population to the dust heating in M\,51 is consistent with the upper limit found by \citet{2000A&A...362..138P,2001A&A...372..775M} for five edge-on spiral galaxies. \citet{2011ApJ...738..124L} report an even lower contribution with less than 20$\%$ of energy to the global dust heating provided by old stars. The importance of young stars contributing to the heating of the diffuse, cold dust component is also evident from the tight correlation between the effective SFR (SFR/$L_{\text{500}}$) and the cold dust temperature ($T_{\text{c}}$) for 20 local star forming galaxies of the Key Insights on Nearby Galaxies: A Far-IR Survey with \textit{Herschel} (KINGFISH) sample \citep{2014ApJ...789..130K}. The scatter in the latter relation and a trend between $L_{\text{3.6}}$/$L_{\text{500}}$ and the dust emissivity index $\beta$, however, points at a non-negligible (although non-dominant) contribution from the old stellar population to the dust heating at FIR/submm wavelengths \citep{2014ApJ...789..130K}.

For the entire dust energy budget emitted between 8 and 1000\,$\mu$m, the dust heating analysis is not affected by non-linear affects and the total-infrared emission by dust heated through young and old stellar photons simply adds up to the total-infrared dust emission.
We find that about 63\,$\%$ of the dust emission is provided through heating by young stars while the remaining 37\,$\%$ is heated through the more evolved stellar population. With M\,51 being located along the main sequence of star-forming galaxies in the local Universe \citep{2007ApJ...660L..43N}, we might need to consider a significant contribution of dust heating from evolved stars for nearby spiral galaxies with similar loci in the SFR-vs-stellar mass diagram (or thus similar sSFR).

\begin{table}
\caption{Dust heating fractions in several monochromatic infrared wavebands $X$, with $F_{\text{X,old}}$ and $F_{\text{X,young}}$ indicating the fraction of energy for radiative dust heating provided by old and young stars ($<$ 100 Myr), respectively.}
\label{table_dustheating}
\centering
\begin{tabular}{lcc}
\hline \hline 
$\lambda$ [$\mu$m] & $F_{\text{X,old}}$  [$\%$] & $F_{\text{X,young}}$  [$\%$] \\
\hline 
24  & 6 & 94 \\
70 & 34  & 66  \\
100 & 39 & 61 \\
160 & 40 & 60 \\
250 & 41 & 59 \\
350 & 42 & 58 \\
500 & 42 & 58 \\
1000 & 41 & 59 \\
\hline
TIR & 37 & 63 \\
\hline 
\end{tabular}
\end{table}

\subsection{Spatially resolved heating of the total-infrared dust emission}
\begin{figure}
\centering
\includegraphics[width=8.5cm]{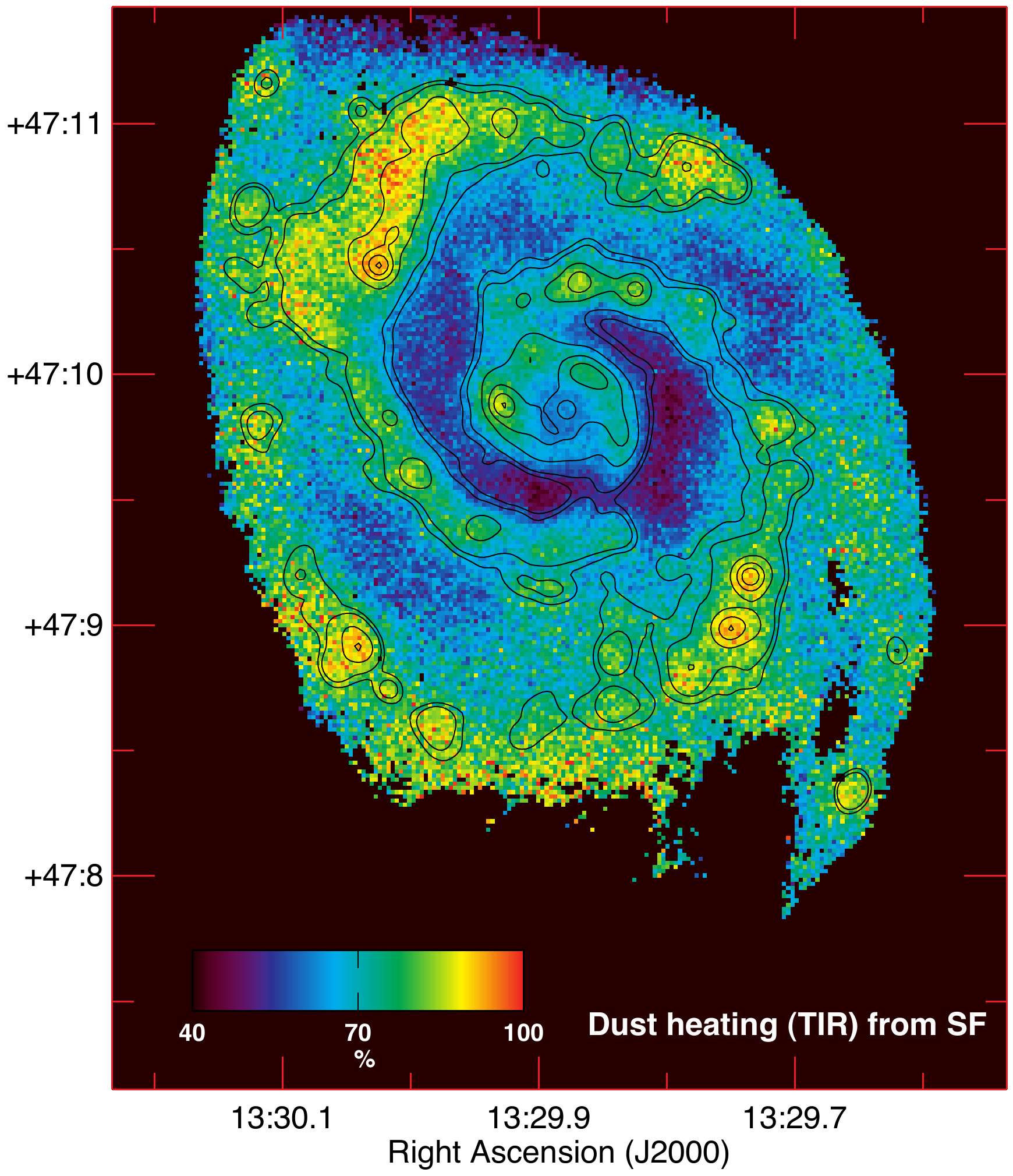}   
 \caption{The contribution from young stars ($\lesssim$ 100 Myr) to the radiative dust heating responsible for the total-infrared emission of M\,51. The black contours indicate the SFR obtained from a combination of FUV+24\,$\mu$m SFR calibrators \citep{2011ApJ...741..124H} with contour levels of 0.075, 0.1, 0.25, 0.5, 0.75, 1.0 $\times$ $10^{-3}$ M$_{\odot}$ yr$^{-1}$.} \label{plot_step2_localdustheating}
\end{figure}

Figure \ref{plot_step2_localdustheating} shows the spatially resolved contribution to the dust heating from star formation to the total infrared emission ranging from about 40$\%$ to 100$\%$. More specifically, we compute the fraction $F'_{\text{TIR,young}}$ of the TIR emission that is due to heating by young stars. In the inter-arm regions of M\,51, the dust heating appears mostly powered by evolved stars. With the black contours indicating the observed SFR in M\,51, we can clearly see that the dust heating follows the morphology of star-forming regions located throughout the spiral arms. Especially, the northern spiral arm region located near the interacting galaxy NGC\,5195 seems dominated by young stellar photons that heat the dust in the spiral arms and well beyond. The peak of $FUV$ emission in this part of M\,51 indeed suggests that star formation has been triggered in the outer regions of the spiral arms in M\,51 with many $UV$ photons capable of escaping from their birth clouds heating up the surrounding dust. 

\subsection{Implications of our dust heating analysis}
The non-negligible contribution from the older stellar population to the heating of dust emitting at infrared/submillimeter wavebands has been suggested in the past based on similar dust energy balance studies and/or optical extinction studies \citep{2000ApJ...539..718C,2000ApJ...528..799W,2008A&A...490..461B, 2008MNRAS.386.1157C,2010A&A...518L..65B,2011AJ....142..111B,2011A&A...527A.109P,2012MNRAS.419.1833B,2012MNRAS.427.2797D} and will have important implications for several extra-galactic studies of the ISM properties in galaxies. More specifically, we caution the applicability of the total infrared emission observed in galaxies as star formation rate tracer given that a significant fraction of the total infrared emission might not be entirely related to the recent star formation activity - at least for galaxies with low to moderate levels of star formation. The main driver of the total infrared energy budget in dusty starburst galaxies might, however, merely be attributed to the young stellar population. 

Besides the questionable usability of $L_{\textit{TIR}}$ as reliable star formation rate diagnostic, the luminosities of far-infrared cooling lines are often compared to $L_{\textit{TIR}}$ (or $L_{\textit{FIR}}$, 42.5-122 or 40-500\,$\mu$m) to trace the photo-electric efficiency in the neutral gas phase, with the cooling lines tracing the gas heating and $L_{\textit{TIR}}$ probing the total dust heating. While the photo-electric effect on dust grains requires the energetic photons of young stars, the line-to-continuum ratios will be diminished by a possible contribution from the old stellar population to the continuum radiation (e.g. \citealt{2013ApJ...776...65P}). Similarly, the total observed $L_{\textit{TIR}}$ of galaxies is often thought to entirely originate from star-forming regions and used to constrain the input parameters of PDR models -neglecting a possible diffuse dust component heated by old stars but inefficient at heating the gas- which will result in an incorrect characterization of the PDR conditions as represented by the scaling factor of the interstellar radiation field, $G_{0}$, and the gas density, $n_{\text{H}}$. 

\subsection{Dust heating prescriptions}
Based on our radiative transfer model for M\,51, we are able to quantify the contribution from different stellar populations to the dust heating at every location in 3D space  in several infrared wavebands. To characterize the dust heating mechanisms, we look at the relation between the dust heating fraction of SF, i.e. the fraction of the energy that is provided by young stars ($<$ 100 Myr) to heat the dust component emitting at IR wavelengths, and the specific star formation rate (sSFR), i.e. the ratio of the current SFR divided by the stellar mass which have been shown to be correlated to the FIR/submm colors of galaxies (e.g. \citealt{2010A&A...518L..61B,2011AJ....142..111B,2012A&A...540A..54B,2014MNRAS.440..942C}). Since the sSFR traces the hardness of the $UV$ radiation field, it is sensitive to the different heating sources in galaxies and, thus, linked to the dust mass fraction heated by star formation \citep{2014A&A...565A.128C}. 

We calculate the specific star formation rate from the output SFR map of the RT calculations, combined with the stellar mass derived from the 3.6\,$\mu$m SKIRT model image adopting the conversion factor to stellar mass reported by \citet{2010MNRAS.405.2279O}. 
Maps of the sSFR and dust heating fraction have been rebinned to a grid with individual pixel sizes of 12.1$\arcsec$, which corresponds to the resolution of the input images to the RT model of M\,51 and, thus, the physical scale on which we can be confident of sampling independent regions. The image pixel size, however, does not guarantee that some of the stars heat the dust in neighboring pixels. 

\begin{figure}
\centering
\includegraphics[width=8.5cm]{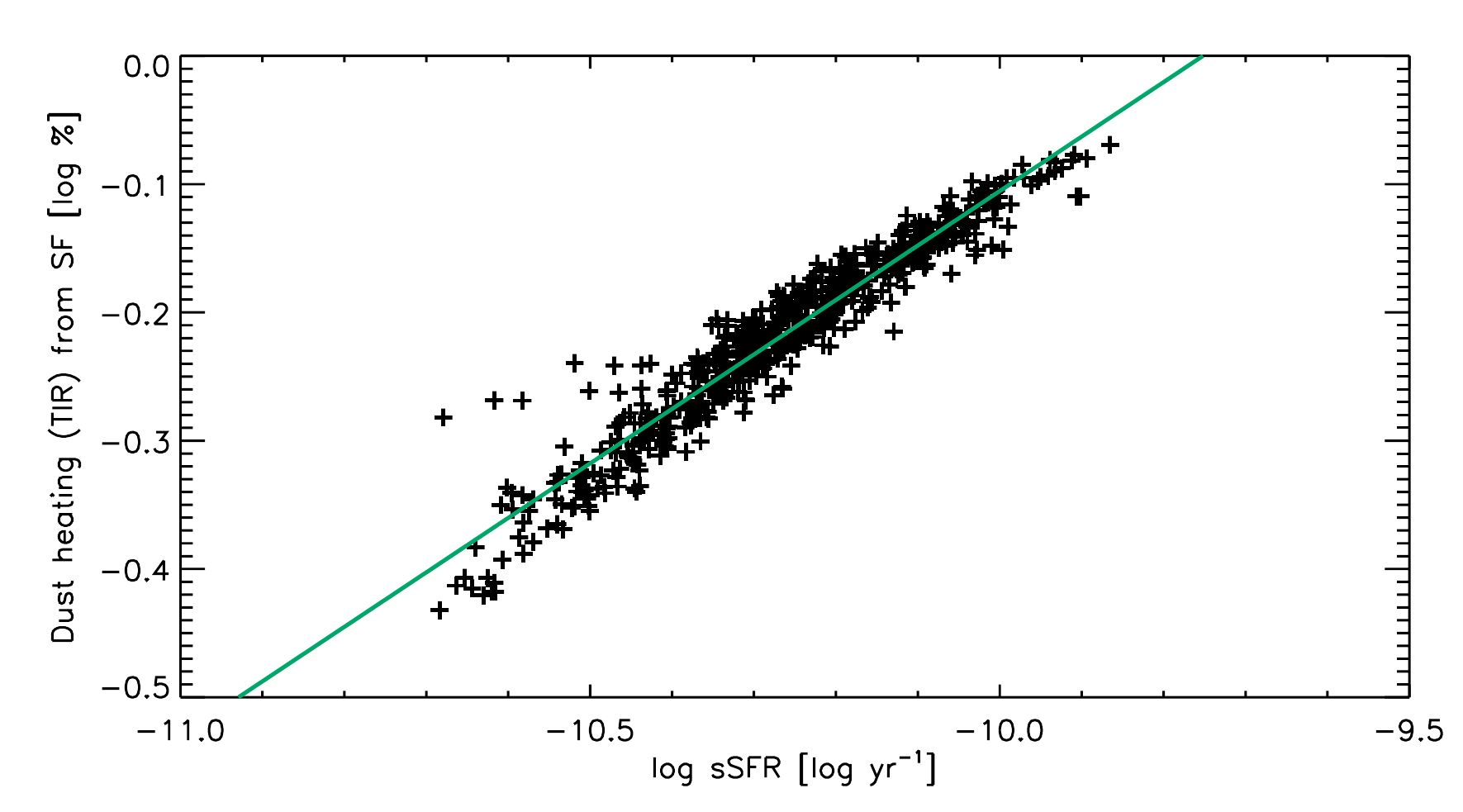}   
 \caption{The relation between the specific star formation rate and the relative contribution from young stars ($\lesssim$ 100 Myr) to the radiative dust heating responsible for the total-infrared emission in the RT model of M\,51. The best fitting line through the observed data points is indicated in green.}
             \label{plot_birthrate}
\end{figure}

Figure \ref{plot_birthrate} shows a tight relation between the sSFR and the fraction of TIR dust heating that can be attributed to star formation (i.e. the relative energy input of young stars to the dust heating). With a correlation coefficient\footnote{The Spearman's rank correlation coefficient, $\rho$, is computed from the IDL procedure r$\_$correlate. Values of $\rho$ close to +1 and -1 are indicative of a strong correlation or anti-correlation, respectively, while values approaching 0 imply the absence of any correlation.} of $\rho$ = 0.95, the importance of star formation in regulating the TIR emission in M\,51 is clearly shown. A similar plot of the relative dust heating fraction \textbf{governed by old stars} as a function of stellar mass (not shown here) does not reveal any particular trend with a correlation coefficient $\rho$ = 0.05, suggesting that the dust heating in M\,51 is dominated by young stars as opposed to the old stellar population.

The best fitting line is fit to the data points in Figure \ref{plot_birthrate} to evaluate the mean trend between the sSFR and the TIR dust heating fraction, i.e. $\log$ dust heating (SF) = 0.42 $\times$ $\log$ sSFR + 4.14. Extrapolating the observed trend for M\,51, we deduce that the total infrared dust heating will be driven solely by young stellar sources for specific star formation rates $\log$ sSFR $\gtrsim$ -9.75 (or SFR $\gtrsim$ 0.2 M$_{\odot}$ yr$^{-1}$, 1.8 M$_{\odot}$ yr$^{-1}$ and 18 M$_{\odot}$ yr$^{-1}$ for galaxies with stellar masses of $10^9$, 10$^{10}$ and 10$^{11}$ $M_{\odot}$, respectively). More than half of the total infrared dust heating could be powered by the main interstellar radiation field for regions with log sSFR $\lesssim$ -10.46 (or SFR $\lesssim$ 0.03 M$_{\odot}$ yr$^{-1}$, 0.3 M$_{\odot}$ yr$^{-1}$ and 3 M$_{\odot}$ yr$^{-1}$ for galaxies with stellar masses of $10^9$, 10$^{10}$ and 10$^{11}$ $M_{\odot}$, respectively). Although the extrapolation of the dust heating mechanisms in M\,51 seems to guarantee that the dust heating in actively star-forming galaxies and starbursts is driven by the young stellar population, we need to be cautious to interpret the TIR emission from more quiescent galaxies with less vigorous star formation activity.

Since the dust heating conditions might differ from one galaxy to another -depending on metallicity, relative dust-star geometry, other heating mechanisms unrelated to SF (e.g. AGN, shocks), we will -in future work- extend the dust heating analysis to a sample of star-forming galaxies representative for the local Universe to get a better handle on the link between the sSFR and the importance of SF to the dust heating processes in galaxies. It will also be of specific interest to compare the dust heating fraction derived from the radiative transfer modeling to the fractions from diffuse versus clumpy media derived from multi-component dust models and SED fitting tools (e.g. \citealt{2007ApJ...663..866D,2009A&A...507.1793N,2010MNRAS.403.1894D}) and empirical color analyses (e.g. \citealt{2010A&A...518L..65B,2010A&A...518L..55G,2011AJ....142..111B,2012MNRAS.419.1833B,2014A&A...565A...4H,Bendo}).

\section{Conclusions}
\label{Con.sec}
We present a panchromatic radiative transfer analysis of the grand-design spiral galaxy M\,51, that accounts for the absorption and scattering processes of stellar light by dust as well as the thermal re-emission of dust in infrared wavebands.
To reproduce the 3D asymmetric structures of stars and dust in M\,51, we rely on a multi-wavelength data set of observations to constrain the geometrical distribution of stars and dust in our RT model. De-projection of these observed morphologies in combination with the assumption of an exponential distribution in the vertical direction defines the 3D structure of stars. The 3D dust distribution is characterized by the $FUV$ dust attenuation map, $A_{\text{FUV}}$, derived from the $TIR$-to-$FUV$ ratio and following the prescriptions from \citet{2008MNRAS.386.1157C}, assuming an exponential $z$ distribution. 

Contrary to the dust energy balance problem encountered in RT models of edge-on spiral galaxies, the RT model for M\,51 is capable of simultaneously explaining the UV/optical dust extinction and far-infrared/sub-millimeter thermal re-emission. We argue that the discrepancy in the dust energy balance in highly inclined systems is mainly related to the inability to probe the dust attenuating effects from UV/optical data and the ignorance on the position of star-forming regions along the line-of-sight. 

We benefit from the high-resolution 3D radiative transfer model -simulating the interaction of stellar light of various energies with the residing dust at every location in 3D space- to get insight in the different dust heating mechanisms that contribute to the dust emission in infrared wavebands. On global scales, we find that the young stars account for the heating of dust responsible for 63$\%$ of the total infrared emission (8-1000\,$\mu$m) while 37$\%$ is governed through heating by the evolved stellar population. In individual wavebands, the contribution from young stars to the dust heating dominates at \textit{ALL} infrared wavebands but gradually decreases towards longer infrared and submillimeter wavebands for which the old stellar population becomes a non-negligible source of heating. The tight correlation between the specific star formation rate (sSFR) and the TIR dust heating fraction from young stars confirms that young stars dominate the dust heating in M\,51.
Upon extrapolation of the observed trends for M\,51, we conclude that the total infrared dust heating will be driven solely by young stellar sources for specific star formation rates $\log$ sSFR $\gtrsim$ -9.75, while more than half of the total infrared dust heating could be powered by the main interstellar radiation field for regions with log sSFR $\lesssim$ -10.46. 

The first application of a new high-resolution 3D RT modeling technique reveals the importance of old stars contributing to the dust emission at all IR wavebands, which will affect the reliability of the total-infrared emission to trace the star formation rate in galaxies with moderate star formation activity. The 3D RT modeling technique presented in this work will be exploited in the future to derive possible variations in dust heating sources, dust geometry and dust composition among different galaxy populations in the local Universe.

\begin{acknowledgements}
We thank Richard Tuffs, Cristina Popescu, Simone Bianchi, Manolis Xilouris and Giovanni Natale for fruitful discussions on the analysis of the dust heating fractions and the clumpiness of dust. We thank Robert C. Kennicutt Jr. for useful suggestions. We also thank Brent Groves for kindly providing us the dust masses associated with the emission spectra of the young H{\sc{ii}} regions with surrounding PDR envelopes. \\
IDL is a postdoctoral researcher of the FWO-Vlaanderen (Belgium).
MB, JF and TH acknowledge the financial support of the Belgian Science Policy Office (BELSPO) through the PRODEX project $"$Herschel-PACS Guaranteed Time and Open Time Programs: Science Exploitation$''$ (C90370).
PC acknowledges the financial support of the Belgian Science Policy Office (BELSPO) through the CHARM framework (Contemporary physical challenges in
Heliospheric and AstRophysical Models), a phase VII Interuniversity Attraction Pole (IAP) programme.
LC acknowledges support under the Australian Research Council's Discovery Projects funding scheme (project number 130100664).
PACS has been developed by a consortium of institutes
led by MPE (Germany) and including UVIE
(Austria); KU Leuven, CSL, IMEC (Belgium);
CEA, LAM (France); MPIA (Germany); INAFIFSI/
OAA/OAP/OAT, LENS, SISSA (Italy);
IAC (Spain). This development has been supported
by the funding agencies BMVIT (Austria),
ESA-PRODEX (Belgium), CEA/CNES (France),
DLR (Germany), ASI/INAF (Italy), and CICYT/
MCYT (Spain). SPIRE has been developed
by a consortium of institutes led by Cardiff
University (UK) and including Univ. Lethbridge
(Canada); NAOC (China); CEA, LAM
(France); IFSI, Univ. Padua (Italy); IAC (Spain);
Stockholm Observatory (Sweden); Imperial College
London, RAL, UCL-MSSL, UKATC, Univ.
Sussex (UK); and Caltech, JPL, NHSC, Univ.
Colorado (USA). This development has been
supported by national funding agencies: CSA
(Canada); NAOC (China); CEA, CNES, CNRS
(France); ASI (Italy); MCINN (Spain); SNSB
(Sweden); STFC and UKSA (UK); and NASA
(USA).
\end{acknowledgements}

\begin{appendix} 

\section{The effect of model degeneracies}
\subsection{Relative dust-stellar scale heights}
\label{effect_scaleheight}

\begin{figure*}
\centering
\includegraphics[width=18.5cm]{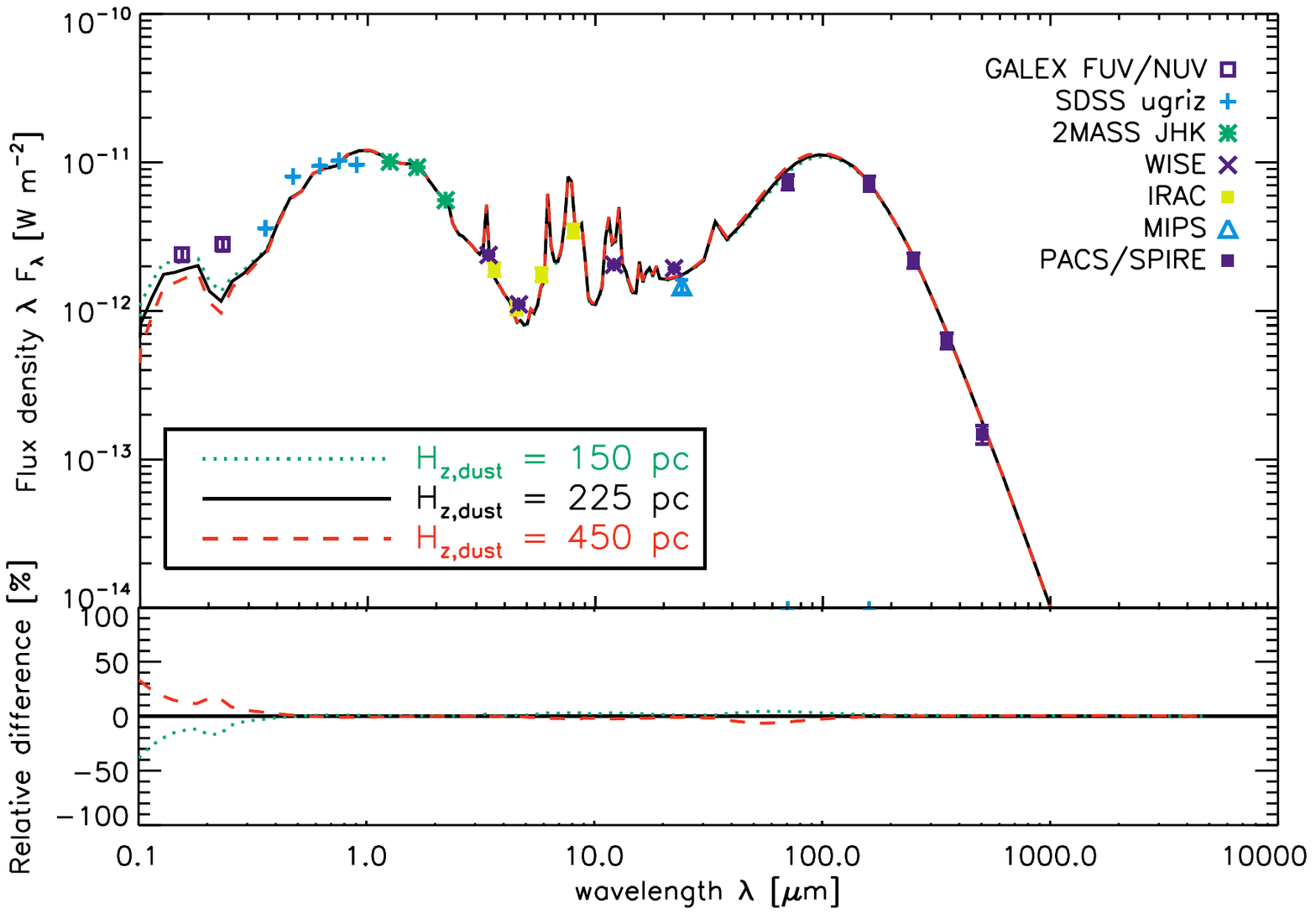}   
 \caption{The effect on the model SED (top) derived from the RT simulations for M\,51 assuming a dust scale height $h_{\text{z,d}}$ = 150, 225 and 450 pc or, thus, a dust-to-stellar scale height of $\frac{1}{3}$, $\frac{1}{2}$ and $1$, indicated as a black solid, green dotted and red dashed line, respectively. The bottom panel shows the relative SEDs for RT models with dust scale heights of $h_{\text{z,d}}$ = 150 pc (green dotted line) and $h_{\text{z,d}}$ = 450 pc (red dashed line) compared to the RT model with $h_{\text{z,d}}$ = 225 pc, which was used for the analysis in this paper.}
             \label{plot_compare_sed_hzdust}
\end{figure*}

In Section \ref{Geometry.sec}, we describe our model assumptions for the relative star-dust geometry in M\,51 based on observations of edge-on galaxies for which the stellar and dust geometries can be vertically resolved.
Since the relative geometry of stars and dust has been shown to affect the derived attenuation (e.g. \citealt{1989MNRAS.239..939D,1999ApJS..123..437F}), we aim to analyze the model implications regarding our assumptions on the relative dust-star geometry.
Figure \ref{plot_compare_sed_hzdust} shows the model SEDs (top) and relative difference between the model SEDs (bottom) for a RT model of M\,51 with dust scale height $h_{\text{z,d}}$ = 150, 225 and 450 pc or a dust-to-stellar scale height of $\frac{1}{3}$, $\frac{1}{2}$ and $1$, respectively.

The difference between the three RT models with different dust scale heights is limited to the $UV$ and far-infrared wavelength domain, while the optical wavebands do not reveal a clear dissimilarity between the observed SEDs. The $FUV$ emission decreases for RT models with increasing dust scale heights, i.e. dust scale heights that approach the vertical scale height of stars, suggesting that more $FUV$ emission is absorbed for dust components distributed in a thicker disk. 
Comparing the three models, the $UV$ absorption increases from $\sim$ 75$\%$ to 78$\%$ and 81$\%$ for dust scale heights of $h_{\text{z,d}}$ = 150, 225 and 450 pc, respectively. The scattered fraction of $UV$ light, on the other hand, slightly decreases from 12$\%$ to 11$\%$. The total amount of absorbed stellar energy (i.e. 10$\%$ of the total stellar bolometric luminosity) remains invariable for the three RT models, which confirms earlier results reported by \citet{1996ApJ...467..175B} and \citet{2004A&A...419..821T} that the relative dust-star geometry does not severely affect the attenuation (except in $UV$ wavebands).
The increased absorption of $UV$ light for RT models with longer dust scale heights translates in brighter PAH features and increased mid-infrared emission. The differences between the three RT models with variation in dust scale heights at infrared wavebands are limited to $\sim$ 10$\%$ and, therefore, will not have a significant effect on the RT model parameters. 

\begin{figure*}
\centering
\includegraphics[width=18.5cm]{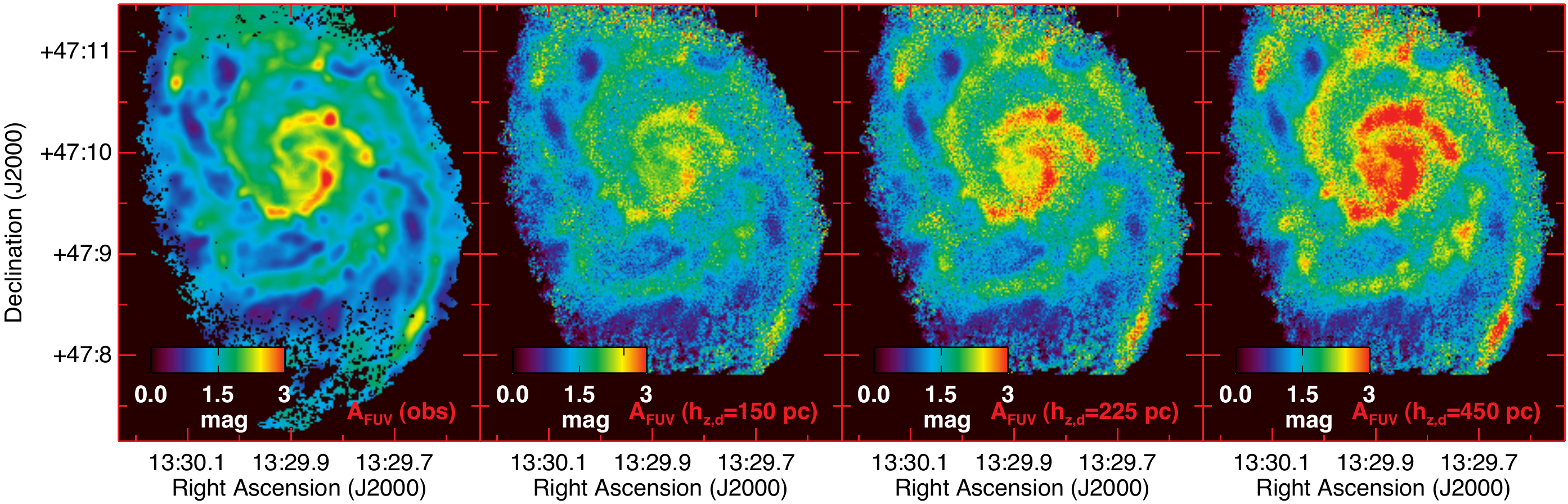}   
 \caption{The $FUV$ attenuation map for M\,51 as derived from observations (left), for the RT model exploited in this paper with dust scale height $h_{\text{z,d}}$ = 150 pc (second panel), $h_{\text{z,d}}$ = 225 pc (third panel) and $h_{\text{z,d}}$ = 450 pc (last panel).}
             \label{plot_compare_afuv_hzdust}
\end{figure*}

Figure \ref{plot_compare_afuv_hzdust} shows the observed (first panel), and model images of the $FUV$ dust attenuation in M\,51 for RT models with dust scale height $h_{\text{z,d}}$ = 150 pc (second panel), $h_{\text{z,d}}$ = 225 pc (third panel) and $h_{\text{z,d}}$ = 450 pc (last panel).
From the $FUV$ attenuation maps, we can derive that the absorbed $FUV$ fraction is indeed higher/lower for larger/smaller dust scale heights. Since an adjustment of the dust mass would be necessary to reproduce the observed $FUV$ attenuation map for RT models with dust scale heights  $h_{\text{z,d}}$ = 150 and 450 pc, respectively, we will not longer reproduce the observed far-infrared/sub-millimeter observations of M\,51. We, therefore, argue that the assumed relative dust-to-stellar scale height $h_{\text{z,d}}$/$h_{\text{z,}\star}$ = 0.5 provides a good representation of the true vertical distribution of stars and dust in M\,51. We do need to be cautious regarding uncertainties on the assumed dust model parameters, i.e. deviations from the assumed dust opacity $\kappa$ and its variation with wavelength (e.g. \citealt{2014A&A...561A..95T,2014arXiv1406.6066G}) will reflect in a similar uncertainty on the dust mass determination, FUV attenuation and dust emission in IR/submm wavebands.

\subsection{Clumpiness of dust}
\label{effect_clumps}

The RT model of M\,51 does not include any effects of clumpy dust in addition to the 2D asymmetry derived from the $FUV$ attenuation map to describe the dust geometry. 
To analyze the effects of dust clumping on the RT model results for M\,51, we introduce a two-phase geometry in our radiative transfer code SKIRT characterized by a diffuse and clumpy phase.
The two-phase component is conditioned through three parameters: the number of clumps, the size of individual clumps and the clump fraction, i.e. the fraction of the dust mass that is locked up in clumps with the remaining mass distributed in the diffuse medium. Rather than using a density ratio between the clumps and diffuse medium, we model the clumpy regions using a Cubic Spline Smoothing kernel $W(\textbf{r},h)$ to describe the density of the two-phase medium:
\begin{equation}
\rho (\textbf{r})~=~(1-f_{\text{cl}}) \rho_{\text{orig}}(\textbf{r}) + \frac{f_{\text{cl}}}{N} \sum_{i=1}^{N} W(\textbf{r}-\textbf{r}_{i},h)
\end{equation}
with $N =$ the number of clumps, $\rho_{\text{orig}} =$ the original density, $f_{\text{cl}} =$ the fraction of mass locked up in clumps, $\textbf{r} =$ the 3D position in the dust grid and $h =$ the scale radius of a single clump.
The clumps with their centers at position $\textbf{r}_{i}$ are distributed stochastically throughout the diffuse medium relying on the probability density distribution function representative of the dust component.

Similar to the two-phase clumpy medium applied by \citet{2008A&A...490..461B}, the clumps in our model are assumed to have sizes similar to the Giant Molecular Clouds (GMCs) observed in our Milky Way. Based on the largest cloud mass of 10$^{6.5}$ M$_{\odot}$ \citep{2005PASP..117..978R} and assuming a typical surface density $\Sigma$ = 100 M$_{\odot}$ pc$^{-2}$ observed in Local Group galaxies \citep{2003ApJ...599..258R,2007prpl.conf...81B}, we derive a corresponding radius of $R \sim 100$ pc. The number of dust clumps of a single mass depends on the gas-to-dust ratio (G/D) and the clump fraction $f_{\text{cl}}$ through:
\begin{equation}
(G/D)^{-1} \sum_{i=1}^{N_{\text{c}}} M_{\text{i}} = f_{\text{cl}} M_{\text{dust}}
\end{equation}
The gas-to-dust ratio is assumed to be around $\sim$ 105, following the calculations from \citet{2012ApJ...755..165M}.

\begin{figure*}
\centering
\includegraphics[width=18.5cm]{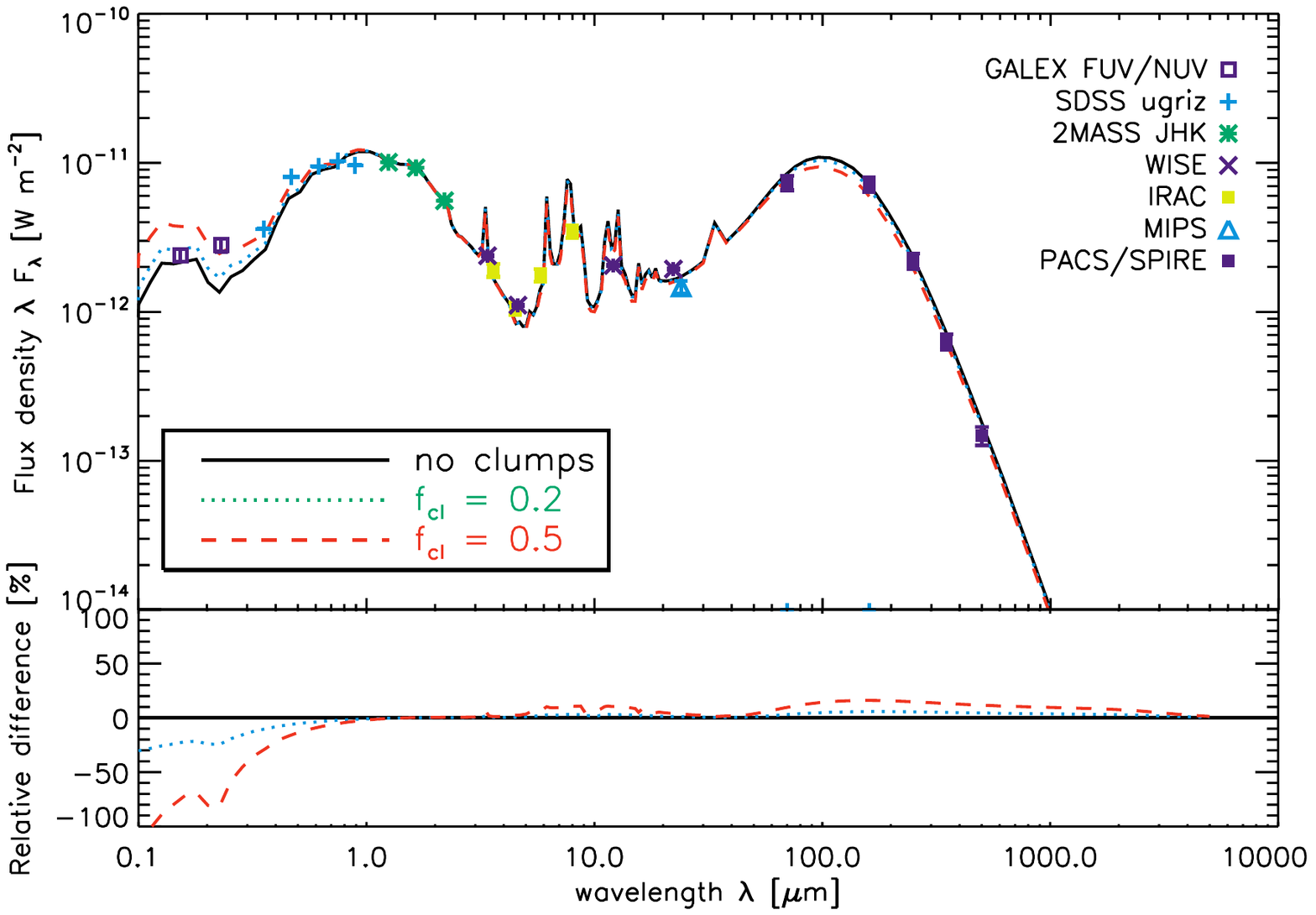}   
 \caption{The effect on the model SED (top) derived from the RT simulations for M\,51 assuming a two-phase dust component with a clump fraction $f_{\text{cl}}$ of 0, 0.2 and 0.5, representative for the relative fraction of dust mass distributed in clumps, indicated as a black solid, green dotted and red dashed line, respectively. The bottom panel shows the relative SEDs for RT models with clump fractions $f_{\text{cl}}$ = 0.2 (green dotted line) and $f_{\text{cl}}$ = 0.5 (red dashed line) compared to the RT model with a smooth dust component, which was used for the analysis in this paper.}
             \label{plot_compare_sed_fcl}
\end{figure*}

Figure \ref{plot_compare_sed_fcl} shows the model SEDs (top) and relative difference between the SEDs (bottom) for radiative transfer simulations with a two-phase dust component consisting of a smooth medium with superimposed a fraction of clumps representing 0$\%$ (equivalent to no clumps), 20$\%$ and 50$\%$ of the total dust mass, indicated in black solid, green dotted and red dashed lines, respectively. With increasing clump fractions, the dusty medium becomes more transparent, resulting in increased $UV$ and optical emission of stars in the RT model. The total amount of stellar light that is absorbed by dust decreases from 10$\%$ to 7$\%$ going from a smooth to clumpy medium with $f_{\text{cl}}$ = 0.5, while the amount of stellar photons that is scattered by the dust medium decreases from 18$\%$ to 15$\%$. The diminished absorption rate is most apparent in the $UV$ wavebands with $UV$ absorption fractions decreasing from 75$\%$ to 55$\%$ going from a smooth to clumpy media with $f_{\text{cl}}$ = 0.5. The $UV$ scattering, on the other hand, increases from about 10$\%$ to 15-20$\%$ from a smooth to clumpy medium with half of the dust mass distributed in clumps. The clumping fraction of dust has, thus, a direct impact on the $UV$ absorption and scattering rates, which also affects the shape of the 2175$\AA$ feature (see bottom panel of Figure \ref{plot_compare_sed_fcl}). In addition to variations in the grain composition and/or size distribution, the clumping of dust might play an important role in explaining the shape of the dust bump at 2175$\AA$.

\begin{figure*}
\centering
\includegraphics[width=18.5cm]{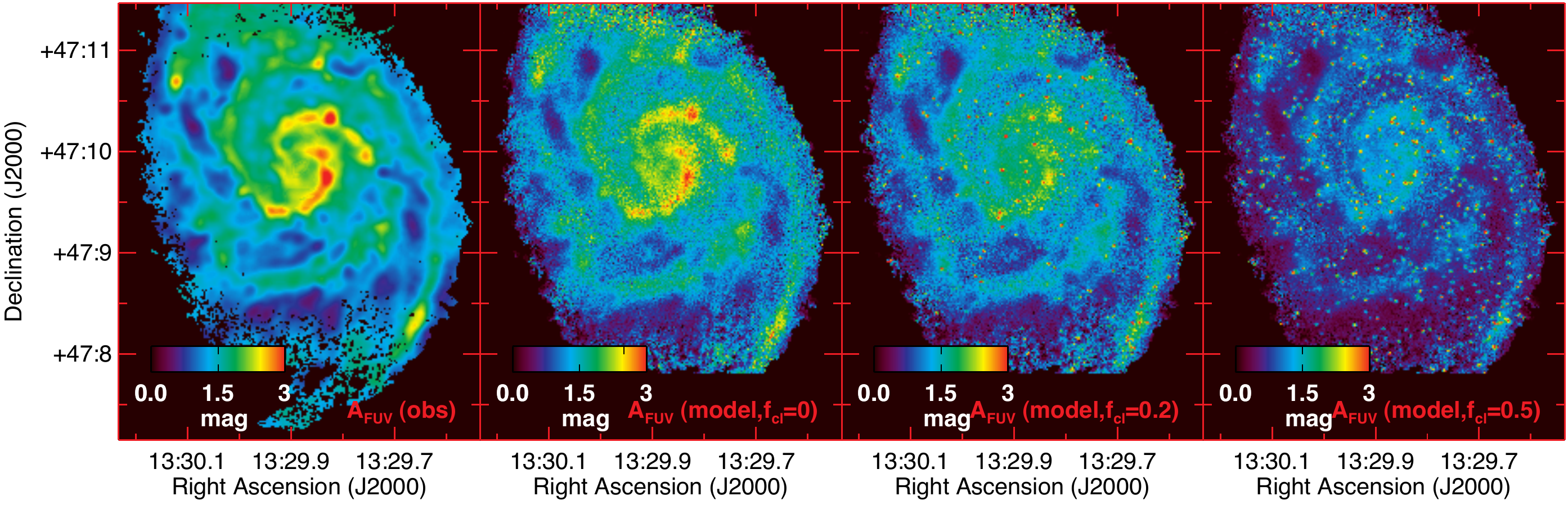}   
 \caption{The $FUV$ attenuation map for M\,51 as derived from observations (left), for the RT model exploited in this paper with smooth dust component (second panel) and the RT model with a two-phase dust component with clump fractions $f_{\text{cl}}$ = 0.2 (third panel) and $f_{\text{cl}}$ = 0.5 (last panel).}
             \label{plot_compare_afuv_fcl}
\end{figure*}

The transparency of the ISM in clumpy media reduces the absorbed stellar energy, which is reflected in a decrease of the infrared and (sub-)millimeter emission with about $\sim$ 10-15$\%$ long wards of 60\,$\mu$m. Similar effects of dust clumping have been shown in RT simulations presented by \citet{2000MNRAS.311..601B,2004ApJ...617.1022P,2011ApJS..196...22B}. The decreased dust emission could be compensated for by increasing the dust mass in the RT model, but would still underestimate the $FUV$ attenuation derived for M\,51 by $\sim$ 40$\%$ (see Figure \ref{plot_compare_afuv_fcl}). We, therefore, argue that the 2D structure from the $FUV$ attenuation map constructed based on $FUV$ and infrared observations provides an adequate description of the 3D asymmetry in the distribution of dust in M\,51. The underestimation of the $FUV$ attenuation upon including a two-phase dust component with clumps suggests that M\,51 likely does not contain many compact dust clouds without embedded sources. This result for M\,51 differs from the substantial fractions of compact dust clumps needed to explain the far-infrared/sub-millimeter emission observed in NGC\,891 ($f_{\text{cl}}$ = 0.5, \citealt{2008A&A...490..461B}), NGC\,4594 ($f_{\text{cl}}$ = 0.75, \citealt{2012MNRAS.419..895D}) and NGC\,4565 ($f_{\text{cl}}$ = 0.67, \citealt{2012MNRAS.427.2797D}). A first difference between these galaxies and M\,51 is their edge-on view, which does not allow to constrain the 3D structure of stars and dust as we did for M\,51 and might, therefore, introduce uncertainties into the model. A second explanation might be the difference in galaxy type between the former galaxies and M\,51. In late-type galaxies such as M\,51 (SAbc), we expect to find star-forming regions with embedding young stars, while the early-type galaxies (Sa-Sb) likely have lower levels of star formation activity with several compact gaseous clouds where star formation has not (yet) been ignited. The study of a bigger sample of face-on galaxies will be necessary to disentangle between the effects of inclination and galaxy type in explaining the need for compact quiescent dust clouds. 

\end{appendix}

\end{document}